\documentclass[conference]{IEEEtran}

\usepackage{graphicx}
\usepackage{subfig}
\usepackage{amsmath}
\usepackage{amsfonts}
\usepackage{booktabs}
\usepackage{multirow}
\usepackage{wrapfig}
\usepackage[colorlinks,
  linkcolor={green!80!black},
  citecolor={blue},
  urlcolor={blue!70!black}]{hyperref}
\usepackage{cleveref}
\usepackage{amsthm}
\usepackage{bbm}
\usepackage{newfloat}
\usepackage{wasysym}
\usepackage{xcolor}
\usepackage{amssymb}
\usepackage{bbm}
\usepackage{tikz}
\usepackage{algorithm}
\usepackage{algorithmicx}
\usepackage{algpseudocode}
\usepackage[verbose=silent]{microtype}
\usepackage {setspace}
\usepackage{enumitem}
\setlist{leftmargin=7mm}
\usepackage{pifont}
\usepackage{amsthm}
\usepackage{colortbl}
\usepackage{array}
\usepackage{threeparttable}
\usepackage[most]{tcolorbox}
\usepackage[framemethod=default]{mdframed}
\newtcolorbox{takeawaybox}{
  colback=gray!20,
  colframe=gray!20,
  coltitle=black,
  arc=4pt,
  boxrule=0.5pt,
  boxsep=2pt,
  left=2pt,
  right=2pt,
  top=2pt,
  bottom=2pt,
  before skip=0.8\baselineskip,
  after skip=0.8\baselineskip
}

\newmdenv[
  linecolor=blue!70,
  linewidth=1pt,
  roundcorner=5pt,
  backgroundcolor=gray!10,
  innertopmargin=\topsep,
  innerbottommargin=\topsep,
  leftmargin=10pt,
  rightmargin=10pt,
]{framedquote}

\theoremstyle{definition}

\newcommand{\partitle}[1]{\vspace{0.3em} \noindent \textbf{#1.}}
\usepackage{makecell}
\usepackage{xspace}

\definecolor{level1}{RGB}{132, 208, 103}
\definecolor{level2}{RGB}{179, 212, 155}
\definecolor{level3}{RGB}{239, 154, 154}
\definecolor{level4}{RGB}{229, 115, 115}
\definecolor{level5}{RGB}{211, 47, 47}

\newcommand{\cellcolorrr}[1]{%
  \ifdim #1pt < 0.25pt \cellcolor{level1!80}#1\else
  \ifdim #1pt < 0.5pt \cellcolor{level2!70}#1\else
  \ifdim #1pt < 0.75pt \cellcolor{level3!70}#1\else
  \cellcolor{level4!85}#1\fi\fi\fi
}

\definecolor{success}{HTML}{A5D6A7}
\definecolor{failure}{HTML}{F48FB1}
\definecolor{warning}{HTML}{FDECB6}

\ifCLASSOPTIONcompsoc
  \usepackage[nocompress]{cite}
\else
  \usepackage{cite}
\fi

\ifCLASSINFOpdf

\else

\fi

\begin{document}

\title{Shadow in the Cache: Unveiling and Mitigating Privacy Risks of KV-cache in LLM Inference}

\author{
\IEEEauthorblockN{Zhifan Luo\textsuperscript{1}, Shuo Shao\textsuperscript{1}, Su Zhang\textsuperscript{2}, Lijing Zhou\textsuperscript{2}, Yuke Hu\textsuperscript{1,*}, Chenxu Zhao\textsuperscript{1}, Zhihao Liu\textsuperscript{1}, Zhan Qin\textsuperscript{1,3,*}
\thanks{*Corresponding author(s).}
}
\IEEEauthorblockA{\textsuperscript{1}State Key Laboratory of Blockchain and Data Security, Zhejiang University  \ \ 
\textsuperscript{2}Huawei Technology \\
\textsuperscript{3}Hangzhou High-Tech Zone (Binjiang) Institute of Blockchain and Data Security\\
\{luozhifan, shaoshuo\_ss, yukehu, zhaocx\_7, zhihao\_liu, qinzhan\}@zju.edu.cn; \{zhangsu14, zhoulijing\}@huawei.com
}
}

\IEEEoverridecommandlockouts
\makeatletter\def\@IEEEpubidpullup{6.5\baselineskip}\makeatother
\IEEEpubid{\parbox{\columnwidth}{
    Network and Distributed System Security (NDSS) Symposium 2026\\
    23-27 February 2026, San Diego, CA, USA\\
    ISBN 979-8-9919276-8-0\\
    https://dx.doi.org/10.14722/ndss.2026.240258\\
    www.ndss-symposium.org
}
\hspace{\columnsep}\makebox[\columnwidth]{}}

\maketitle

\begin{abstract}

The Key-Value (KV) cache, which stores intermediate attention computations (Key and Value pairs) to avoid redundant calculations, is a fundamental mechanism for accelerating Large Language Model (LLM) inference. However, this efficiency optimization introduces significant yet underexplored privacy risks. This paper provides the first comprehensive analysis of these vulnerabilities, demonstrating that an adversary can reconstruct sensitive user inputs directly from the KV-cache. We design and implement three distinct attack vectors: a direct Inversion Attack, a more broadly applicable and potent Collision Attack, and a semantic-based Injection Attack. These methods demonstrate the practicality and severity of KV-cache privacy leakage issues. To mitigate this, we propose KV-Cloak, a novel, lightweight, and efficient defense mechanism. KV-Cloak uses a reversible matrix-based obfuscation scheme, combined with operator fusion, to secure the KV-cache. Our extensive experiments show that KV-Cloak effectively thwarts all proposed attacks, reducing reconstruction quality to random noise. Crucially, it achieves this robust security with virtually no degradation in model accuracy and minimal performance overhead, offering a practical solution for trustworthy LLM deployment.

\end{abstract}

\IEEEpeerreviewmaketitle

\section{Introduction}

Large Language Models (LLMs) have ignited a paradigm revolution in artificial intelligence~\cite{vaswani2017attention, liu2024deepseek}, profoundly impacting various domains and applications, such as machine translation~\cite{zhang2022opt}, chatbots~\cite{yang2024qwen2}, code generation~\cite{he2025benchmarking}, and content creation~\cite{zhao2023survey}. However, the immense scale of these models, characterized by billions or even trillions of parameters, coupled with the need to process extensive input sequences and engage in multi-turn dialogues, presents a substantial challenge to their efficient deployment and inference. This computational demand often translates into high latency and resource consumption, hindering broader accessibility and real-time applicability~\cite{lin2024infinite}.

To address the efficiency bottlenecks in LLM inference, researchers have proposed several optimization techniques~\cite{zhou2024survey, wan2024efficient}. Among these, the \textbf{Key-Value cache (KV-cache)} mechanism has emerged as a crucial and widely adopted solution~\cite{pope2023efficiently, zhao2023survey}. During the autoregressive generation process typical of LLMs, the attention mechanism computes key (K) and value (V) matrices for each token based on its preceding tokens. The KV-cache stores these intermediate attention computations (the K and V pairs) for tokens that have already been processed within the input sequence. By reusing these cached K and V pairs for the generation of subsequent tokens, the KV-cache significantly reduces redundant computations. This dramatically accelerates inference speed and improves throughput, especially for tasks involving long contexts or interactive sessions, making LLMs more practical for real-world deployment.

\begin{figure}[t]
    \centering
    \includegraphics[width=0.89\linewidth]{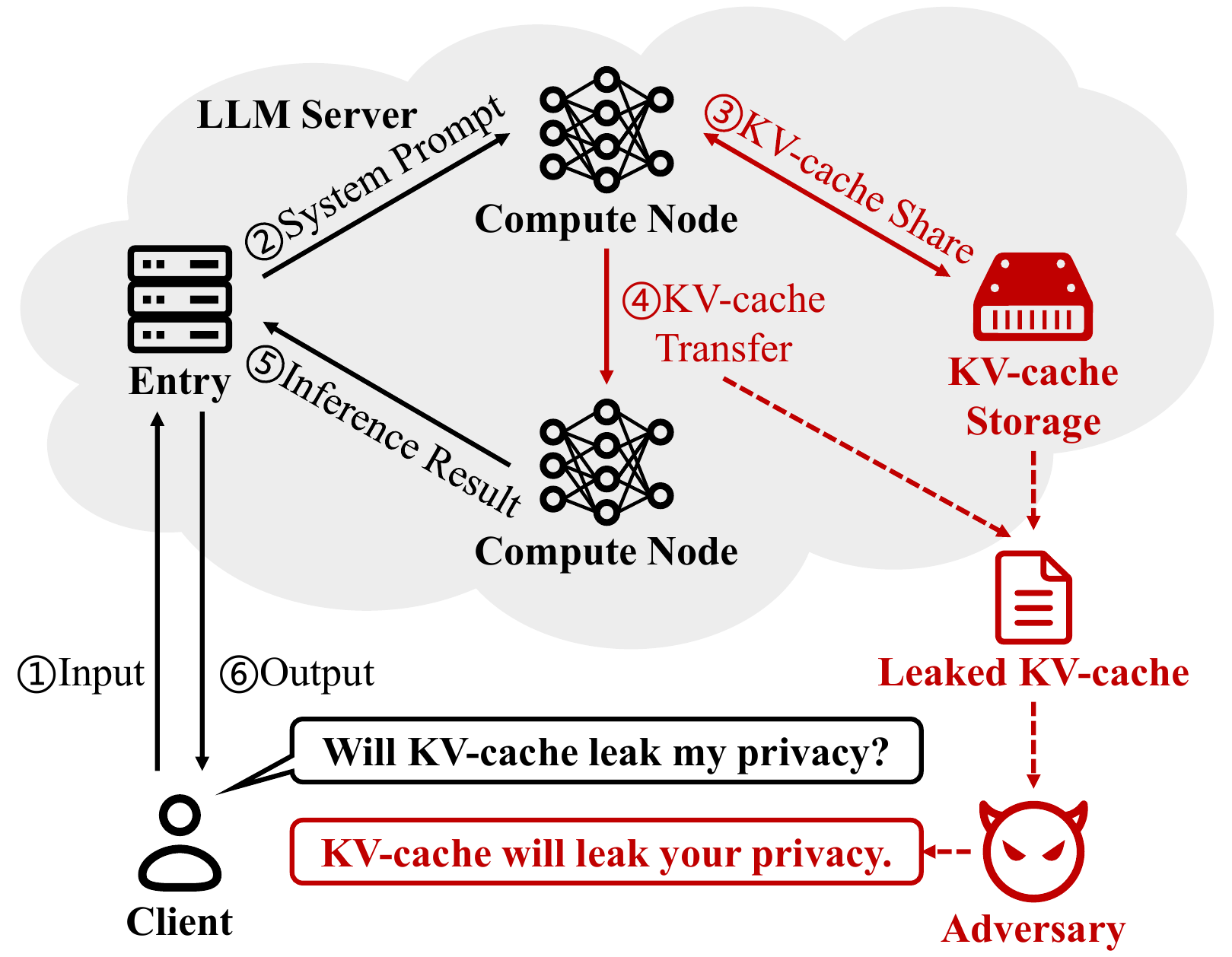}
    \caption{Overview of the privacy-preserving LLM inference workflow and the associated KV-cache leakage threat model. While user-server communication is encrypted (black lines), the KV-cache is often transmitted and stored in plaintext (red dashed lines), creating a surface for privacy attacks.}
    \label{fig:inference_data_leak}
    \vspace{-1.2em}
\end{figure}

However, the storage and potential sharing of the KV-cache introduce significant yet underexplored privacy concerns~\cite{yang2024first,wu2025know}, as illustrated in Figure~\ref{fig:inference_data_leak}. 
This vulnerability stems from a critical trade-off made in production systems: while end-user communication with the LLM service is typically encrypted, the KV-cache itself is almost always processed, transmitted between compute nodes, and persisted in \textbf{plaintext}. 
This design choice is a concession to performance, as the unacceptable latency overhead from cryptographically securing the often gigabyte-scale KV-cache would violate the stringent demands of real-time inference.
Notably, this risk is amplified in the emerging paradigm of confidential Model-as-a-Service (MaaS).
Here, TEE adoption for inference privacy~\cite{apple2024pcc,dhanuskodi2023creating} introduces a critical architectural trade-off: high-throughput architectures require \textit{deliberately externalizing} the massive KV-cache from the TEE's protection boundary.
This architectural design---not a vulnerability---directly exposes the KV-cache to the Cloud Service Provider (CSP), who inherently possesses it without requiring any additional exploit \cite{yuan2025scx}.

Crucially, the direct correlation between the KV-cache and user inputs~\cite{carlini2021extracting, li2025rethinking}, combined with its architectural exposure in plaintext, creates a severe and practical attack vector. 
A critical research gap therefore exists in mitigating these risks, necessitating a practical and secure defense mechanism.

In this paper, to bridge the research gap, we present the first comprehensive study on the privacy risks of KV-cache in LLM inference. Specifically, we primarily study and answer the following two research questions:

\vspace{0.3em}
\begin{itemize}[leftmargin=*]
    \item \textbf{RQ1: Is an adversary able to reconstruct user inputs from the KV-caches?}
\end{itemize}

To address this question, we conduct a systematic investigation, demonstrating that attacks against the KV-cache are not only feasible but also diverse and broadly applicable. We design and implement three distinct classes of targeted privacy-stealing attacks. 
Each attack exposes inherent privacy risks of KV-caches from a different perspective. 

We first explore two relatively direct attack paths. 
The first is the \textbf{KV-cache Inversion Attack}, leveraging known model weight matrices to directly reverse-calculate the input from the KV-cache. 
The second, more general-purpose approach is the \textbf{KV-cache Collision Attack}. 
This attack reframes input reconstruction as a matching task based on forward computation: an adversary iteratively uses a local model instance to generate KV-caches for candidate inputs and compares them against the intercepted target KV-cache to find a match. 
Because this method does not rely on any reverse computation, it has broader applicability. 
Finally, we introduce a novel, semantic-based \textbf{KV-cache Injection Attack}. 
This attack leverages the LLM's powerful capability to understand and execute instructions. 
By appending a specific instruction, such as ``Repeat the previous content,'' to the end of an intercepted KV-cache context, an adversary can induce the model to ``echo'' or generalize the core information held within the KV-cache. 

Collectively, these attacks reveal that privacy leakage from the KV-cache is not merely a theoretical concern. 
Their diversity and feasibility constitute a significant threat to real-world LLM-based applications, underscoring the urgent necessity of designing specialized and efficient privacy-preserving mechanisms for the KV-cache.

\vspace{0.3em}
\begin{itemize}[leftmargin=*]
    \item \textbf{RQ2: Can defenders effectively and efficiently mitigate or prevent user privacy leakage from the KV-cache?}
\end{itemize}

In response to this challenge, this paper provides an affirmative answer by first analyzing the shortcomings of existing privacy-preserving techniques. Conventional methods like full cryptographic encryption~\cite{acar2018survey} or Homomorphic Encryption (HE)~\cite{paillier1999public} introduce prohibitive computational overhead and latency, rendering them impractical for the high-throughput demands of LLM inference. Meanwhile, applying Differential Privacy (DP)~\cite{Dwork2006Calibrating} requires adding a level of noise that often severely degrades the model's inference accuracy to an unacceptable degree. Even specialized solutions like KV-Shield~\cite{yang2024first}, while lightweight, suffer from critical security flaws. Their fixed shuffling mechanism, as we analyzed in Section~\ref{defense:flaws_of_kvshield}, is vulnerable to statistical analysis and is incompatible with modern LLM architectures that use features like Rotary Positional Embedding (RoPE)\cite{su2024roformer}. These limitations highlight the urgent need for a novel defense mechanism.

Therefore, we propose KV-Cloak, a lightweight and secure KV-cache obfuscation mechanism addressing these challenges. 
At its core, KV-Cloak employs a reversible matrix-based obfuscation scheme that guarantees lossless model accuracy. 
Its security is multi-layered: it applies secret invertible linear transformations to obscure statistical properties, and critically, introduces a one-time random permutation matrix for each data block. 
This dynamic permutation prevents adversaries from building stable algebraic relations across multiple queries. 
To further enhance performance, KV-Cloak utilizes operator fusion, algebraically fusing a portion of the secret obfuscation matrices into the LLM's attention layer weights offline.
This shifts the primary computational cost away from the latency-sensitive online inference phase, striking an effective balance between robust security, lossless accuracy, and high performance.

Our contributions can be summarized as follows.

\begin{itemize}[leftmargin=*]
    \item \textbf{Revealing the privacy risks of KV-cache in LLM inference:} We systematically investigate the privacy risks of the KV-cache by designing and implementing three distinct attacks, Inversion, Collision, and Injection attacks.
    \item \textbf{Proposing a lightweight and effective method to mitigate privacy leakage}: We propose KV-Cloak, a novel and practical defense mechanism that uses a lightweight, reversible matrix-based obfuscation scheme combined with operator fusion to protect the KV-cache without degrading model accuracy and with minimal performance overhead.
    \item \textbf{Conducting extensive experiments to evaluate attacks and defenses:} 
    We conduct a systematic evaluation to empirically demonstrate and quantify the feasibility of attacks that reconstruct user input from the KV-cache of state-of-the-art LLMs, establishing it as a practical and severe threat. We further prove that our proposed KV-Cloak is a highly practical solution that achieves robust security, near-lossless model fidelity, and high efficiency simultaneously. Our experiments show that KV-Cloak, with its negligible impact on accuracy and minimal latency overhead (mostly $<1\%$), successfully resolves the stark trade-off between security and utility that plagues alternative approaches such as DP.
\end{itemize}

\section{Background and Related Work}

\subsection{Transformer-based LLM Inference}
Prevailing LLMs, such as LLaMA~\cite{touvron2023llama}, DeepSeek~\cite{liu2024deepseek} and Qwen~\cite{yang2024qwen2}, are predominantly based on the Transformer decoder architecture~\cite{zhao2023survey}. 

\partitle{Self-Attention Mechanism} 
For an input token $x_i$, the self-attention layer generates Query ($q_i$), Key ($k_i$), and Value ($v_i$) vectors via linear transformations. Crucially, modern architectures incorporate RoPE, denoted as ${R}_{\Theta, i}^d$:
\begin{equation}
\label{eq:linear_transform}
    q_{i} = x_{i} W_{q}^{\top} {R}_{\Theta, i}^d, \;\; k_{i} = x_{i} W_{k}^{\top} {R}_{\Theta, i}^d, \;\; v_{i} = x_{i} W_{v}^{\top}.
\end{equation}
where $W_{(\cdot)}$ are learnable weight matrices. The attention score $a_{ij}$ is computed as the scaled dot product between $q_i$ and preceding keys $k_{j \le i}$. The final output $o_i$ is the weighted sum of values projected by $W_o$:
\begin{equation}
\label{eq:attention_score}
    a_{ij} = \frac{\exp(q_{i} {k_{j}}^{\top}/\sqrt{d})}{\sum_{t=1}^{i}\exp(q_{i} {k_{t}}^{\top}/\sqrt{d})}, 
    \;\; o_{i} = \left(\textstyle\sum_{j=1}^{i}a_{ij} v_{j}\right)W_o^\top.
\end{equation}

\partitle{Autoregressive LLM Inference}
LLM inference models the sequence probability $P(x_1, \ldots, x_n)$ through autoregressive decomposition~\cite{bengio2003neural}. 
The generation process predicts tokens sequentially based on the conditional probability:
\begin{equation}
    \label{eq:autoregressive_decomposition}
    P(x_1, \ldots, x_n) = \prod_{i=1}^{n} P(x_i | x_{<i}).
\end{equation}
While effective for capturing dependencies, this sequential nature implies that generating $x_i$ depends on the full history $x_{<i}$. Without caching, the model must re-compute the Key and Value vectors for all preceding tokens at each step. This computational redundancy creates a substantial bottleneck for real-time inference.

\partitle{KV-cache in LLM Inference}
To mitigate the computational redundancy of autoregressive generation, LLMs utilize the \textit{KV-cache} mechanism~\cite{pope2023efficiently}. 
Instead of recalculating the Key ($K$) and Value ($V$) matrices for the entire context window at each step, the system persists these intermediate states. 
When generating a new token $x_i$, the model only computes the current $q_i, k_i, v_i$, performs attention between $q_i$ and the cached history, and subsequently appends the new $(k_i, v_i)$ pair to the cache. 
This strategy significantly reduces latency but necessitates managing a persistent state proportional to the sequence length.

\subsection{Privacy Attacks against LLMs}

\subsubsection{Inference-Phase Inversion Attacks}
Inference-time privacy attacks are categorized into: 
(1) \textbf{Output-based}, reconstructing inputs from final probabilities or embeddings~\cite{li2023sentenceembeddingleaksinformation, morris2023textembeddingsrevealalmost}; 
and (2) \textbf{Intermediate-state-based}, exploiting internal representations~\cite{wan2024information}. 
While early works like Vec2Text~\cite{morris2023textembeddingsrevealalmost} focused on sentence embeddings, Embed Parrot~\cite{wan2024information} demonstrated that deep hidden states allow for more accurate reconstruction.

However, current research often overlooks the \textit{KV-cache}. 
Unlike fused hidden states, the KV-cache retains the primitive Key ($K$) and Value ($V$) vectors—the raw inputs to the attention mechanism. 
Crucially, to support autoregressive generation, the KV-cache is architecturally designed for \textit{persistence}, making it a richer and more exploitable surface than transient computational states.

\subsubsection{Limitations of Side-Channel Attacks}
Existing threats to the KV-cache are primarily side-channel attacks, such as PromptPeek~\cite{wu2025know}, which infers prompts by measuring latency variations caused by cache sharing. 
However, the practicality of such timing attacks is severely mitigated by modern memory management. 

Systems like PagedAttention~\cite{Kwon2023EfficientMM} manage the KV-cache in non-contiguous blocks (defaulting to 16 tokens). 
To trigger a cache hit (and thus a timing signal), an adversary must correctly guess an entire block simultaneously. With a vocabulary size $|V| \approx 10^5$, the search space explodes to $|V|^{16}$, rendering brute-force side-channels computationally infeasible. 
In contrast, we address the \textit{direct access} threat model. 
Given that KV-caches are often processed in plaintext for performance, a compromised system allows adversaries to bypass probabilistic guessing and perform precise, per-token collision attacks.

\section{Attack Landscape: Input Reconstruction from KV-cache}

While the KV-cache serves as a cornerstone for efficient LLM inference, it concurrently introduces a critical and underexplored privacy surface. 
Unlike deep hidden states that represent fused semantic information, entries in the KV-cache maintain a direct, element-wise correspondence to the tokens in the user's input sequence. 
This structural characteristic implies that an adversary gaining access to the KV-cache does not merely obtain abstract embeddings, but potentially holds the raw materials to reverse-engineer the original user prompt.

In this section, we investigate the central research question: \emph{Is an adversary able to reconstruct user inputs from the KV-cache?} 
We first define a realistic threat model grounded in modern cloud architectures, and subsequently detail three distinct attack vectors designed to exploit this vulnerabilities.

\subsection{Threat Model}

This work focuses on the threat landscape of LLM inference services within a confidential computing paradigm.
We consider the inference service provider as the adversary.

\partitle{Adversary's Objective}
The primary objective of the adversary is to recover the user input prompt from the accessed KV-cache. 
We target verbatim reconstruction because it represents the most severe breach of privacy, allowing the extraction of exact credentials, PII, or proprietary logic contained in the input.

\partitle{Adversary's Capabilities}
We assume the adversary can obtain both the \textit{KV-cache} and the \textit{model weights}, but does not observe the ephemeral runtime activations within the GPU registers.
These assumptions are well-grounded in the realities of current confidential computing inference paradigms:
\begin{itemize}[leftmargin=*]
    \item \textbf{KV-cache Access.} The adversary is capable of obtaining the KV-cache stored by the LLM server. 
    This is a realistic assumption, as high-throughput and scalable cloud-native services \textit{deliberately externalize} the large-scale KV-cache to non-secure memory or persistent storage pools to meet performance demands~\cite{yuan2025scx}.
    We thus consider this access a result of an intentional performance-security trade-off, not a traditional vulnerability.
    \item \textbf{Foundation Model Access.} We assume a gray-box setting where the adversary can access the foundation model weights. 
    This access is realistic through two primary paths: 
    (1) For closed-source services, the CSP inherently owns the model and possesses the weights. 
    (2) For services built on open-source models, the provider may be required to disclose the base model for licensing compliance, or the model can be identified using fingerprinting techniques~\cite{pasquini2025llmmapfingerprintinglargelanguage, chen2022teacher}.
\end{itemize}

\subsection{Input Reconstruction Attacks from KV-cache}

\begin{figure*}[t]
    \centering
    \subfloat[KV-cache Inversion Attack.]
    {
        \label{fig:inversion}\includegraphics[width=0.195\textwidth]{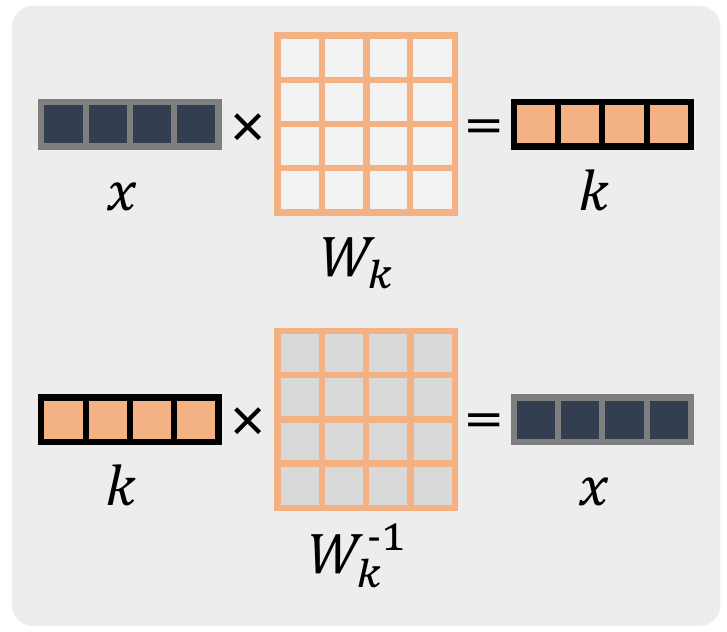}
    }
    \subfloat[KV-cache Collision Attack.]
    {
        \label{fig:collision}\includegraphics[width=0.413\textwidth]{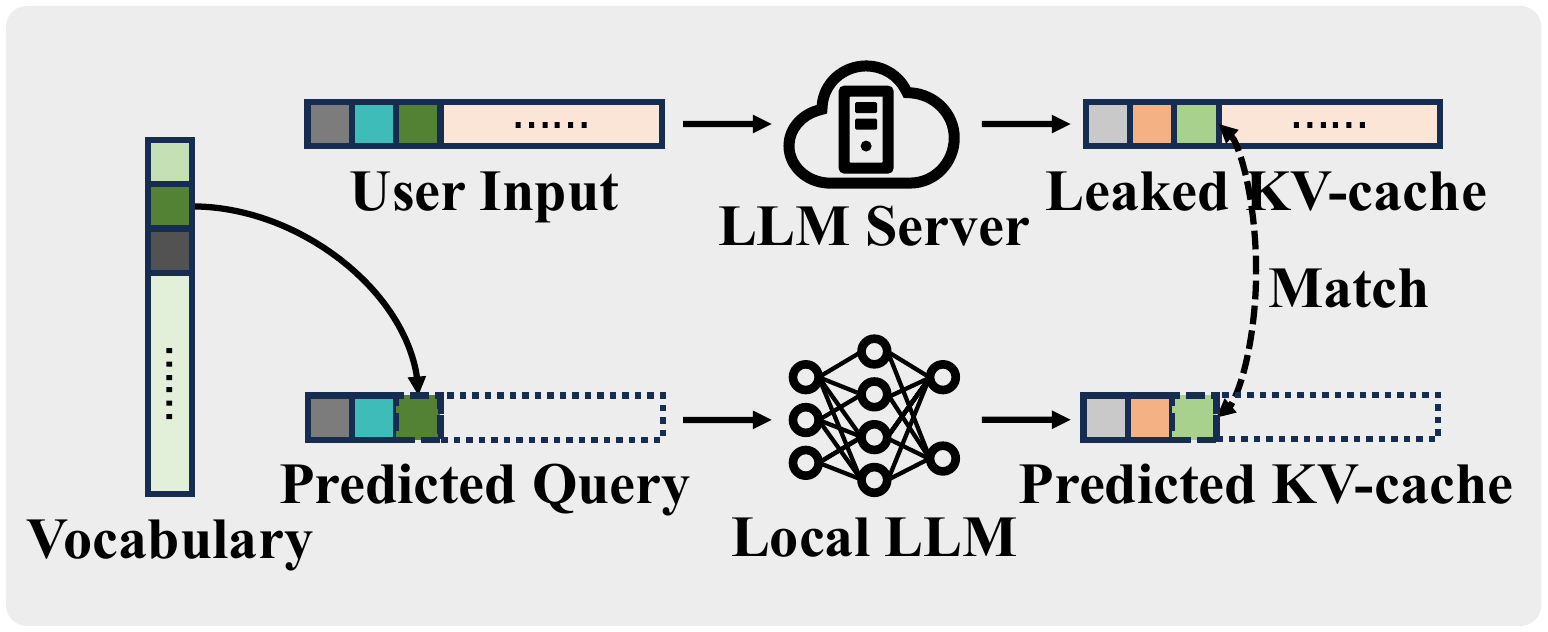}
    } 
    \subfloat[KV-cache Injection Attack.]
    {
        \label{fig:injection}\includegraphics[width=0.35\textwidth]{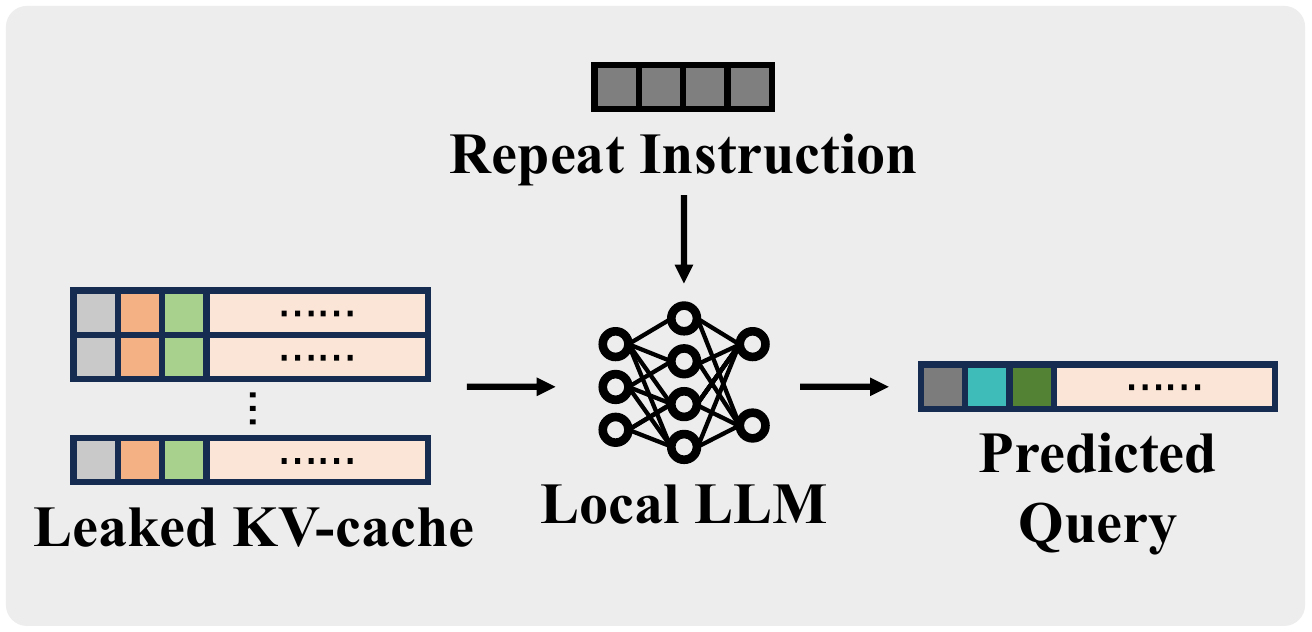}
    }
    \caption{Workflow of the three proposed KV-cache input reconstruction attacks.}
    \label{fig:workflow_of_attacks}
    \vspace{-1.2em}
\end{figure*}

This section investigates the risk of user input leakage from KV-cache data during LLM inference. We systematically present three attack vectors, namely \emph{Inversion Attack}, \emph{Collision Attack}, and \emph{Injection Attack}, for reconstructing the original user inputs from the KV-cache. These attacks differ in complexity, applicability, and exploited vulnerabilities, collectively illustrating the threat landscape.

\subsubsection{KV-cache Inversion Attack}
As illustrated in Figure~\ref{fig:inversion}, the KV-cache Inversion Attack functions as a baseline algebraic adversary. 
It attempts to mathematically invert the attention mechanism by exploiting the linear projection of Key ($k$) and Value ($v$) vectors.

\partitle{Attack Formulation}
Based on the forward pass definition in Eq.~(\ref{eq:linear_transform}), an adversary possessing the plaintext KV-cache and the model weights ($W_k, W_v$) can theoretically recover the input state $x_i$ of the attention layer via matrix inversion:
\begin{equation}
    \label{eq:inversion_attack}
    x_i = k_i ({R}_{\Theta, i}^d)^{-1} (W_k^{\top})^{-1}, \;\; x_i = v_i (W_v^{\top})^{-1}.
\end{equation}

\partitle{Architectural and Semantic Constraints}
While algebraically sound, the feasibility of this attack in production environments is severely limited by two fundamental constraints: 
\textbf{(1)} The attack strictly requires the projection matrices $W_k$ and $W_v$ to be square and full-rank \textbf{invertible}. 
This condition holds for legacy Multi-Head Attention (MHA)~\cite{vaswani2017attention} architectures (e.g., LLaMA-7B). 
However, modern State-of-the-Art (SOTA) models, including LLaMA-3~\cite{grattafiori2024llama}, Qwen~\cite{yang2024qwen2}, and DeepSeek~\cite{liu2024deepseek}, adopt efficiency optimizations such as Grouped-Query Attention (GQA)~\cite{ainslie2023gqa} or Multi-Head Latent Attention (MLA)~\cite{Shao2024DeepSeekV2AS}. 
These mechanisms typically project inputs into lower-dimensional subspaces, resulting in non-square matrices where unique inversion is mathematically impossible (i.e., the problem becomes ill-posed).
\textbf{(2)} the attack's effectiveness is largely confined to the \textbf{first decoder layer's} KV-cache, as its inversion directly yields the input sequence's embeddings, allowing for precise mapping back to the vocabulary. 
In contrast, applying Eq.~(\ref{eq:inversion_attack}) to deeper layers recovers intermediate hidden states. 
These states represent semantically fused contextual information rather than discrete tokens. 
Reversing these deep representations into text is a non-trivial problem that necessitates training auxiliary inversion models~\cite{wan2024information}, thereby increasing the adversary's computational cost and reducing fidelity.

Consequently, while the Inversion Attack serves as a theoretical proof-of-concept for KV-cache leakage, its reliance on specific architectural properties (MHA) and layer positions (Layer 0) limits its utility against modern, deep LLMs. This motivates the need for the more robust attacks.

\subsubsection{KV-cache Collision Attack}
\label{attack:collision}
Unlike the Inversion Attack which suffers from architectural constraints, the \textbf{KV-cache Collision Attack} is a universal, forward-matching adversary applicable to any layer. 
Instead of algebraically reversing the projection, this attack reframes input reconstruction as a search problem: identifying the token in the vocabulary that, when processed by a local model, yields a KV-cache entry most similar to the intercepted target.

\partitle{Fundamental Principle: Reconstruction as Optimization}
The core premise is that the KV-cache exhibits \textit{distance-preserving} properties for identical tokens under the same context. 
We formalize the attack as finding the optimal token $t^*$ at position $i$ that minimizes the distance metric $\mathcal{D}$ (e.g., Frobenius norm) between the leaked cache $K_{\text{leaked}}$ and the local generation $K_{\text{local}}(t)$:
\begin{equation}
    t^*_i = \mathop{\arg\min}_{t \in \mathcal{V}} \mathcal{D}\left(K_{\text{leaked}}^{(i)}, K_{\text{local}}^{(i)}(t | x_{<i})\right).
\end{equation}
The attack's success hinges on \textit{statistical separability}: the distance for the ground-truth token, $d_{\text{target}}$, must be statistically distinguishable from the distribution of distances for incorrect tokens, $d_{\text{other}}$.

\partitle{Attack Procedure}
As illustrated in Figure~\ref{fig:collision}, the attack proceeds iteratively token-by-token. 
For each unknown position $i$: 
\textbf{(1)} \textit{Candidate Generation:} The adversary selects candidate tokens from the model's vocabulary $\mathcal{V}$ to test. 
\textbf{(2)} \textit{Local Simulation:} For each candidate token $t$, the adversary performs a forward pass using their local model instance to generate the corresponding KV-cache entry $K_{\text{local}}^{(i)}(t)$.
\textbf{(3)} \textit{Collision Matching:} The adversary computes the distance $\mathcal{D}$ between the local entry and the intercepted target $K_{\text{leaked}}^{(i)}$. 
The candidate that minimizes this distance (i.e., creates a ``collision'' with the leaked state) is identified as the correct token $x_i$, appended to the sequence, and the process advances to $i+1$.

\partitle{Implementation Optimizations}
A naive exhaustive search over the entire vocabulary ($\mathcal{V} \approx 10^5$) entails prohibitive latency and VRAM consumption, as verifying the \textit{global minimum} distance requires inferring every candidate token. 
To render the attack practical, we implement three synergistic optimizations that transform the problem from a global search into an efficient sequential decision process:
\textbf{(1) Batched Outlier Detection (Enabling Early Exit).} 
Instead of calculating distances for the full vocabulary to find the minimum, we process candidates in batches. 
We fundamentally shift the decision logic: rather than comparing a candidate against all other tokens in $\mathcal{V}$, we determine if a candidate constitutes a \textit{statistical outlier} within its current batch (i.e., $d_t < \mu_{\text{batch}} - 3\sigma_{\text{batch}}$). 
Crucially, this allows for an \textit{early exit}: once a collision is statistically confirmed within a batch, the search terminates immediately without evaluating the remaining vocabulary.
\textbf{(2) Probability-Guided Prioritization.}
To maximize the efficacy of the early exit mechanism, we employ a probability-guided search. 
We sort the candidate tokens based on the model's predicted probabilities $P(x_i|x_{<i})$. 
This reordering ensures that the correct token—which typically carries a high probability—is placed in the \textit{first few batches}. 
Consequently, the collision is triggered near the start of the search process, drastically reducing the average number of inferences required.
\textbf{(3) Search Space Pruning.}
Distinct from prioritization, we explicitly truncate the search space by discarding the ``long tail'' of tokens with negligible probabilities (e.g., keeping only the top-$k\%$ candidates). 
While prioritization accelerates the average case, pruning bounds the \textit{worst-case complexity}. 
It prevents the adversary from wasting resources iterating through tens of thousands of implausible tokens in the rare event that the target is not found in the high-probability regions.

\partitle{Enhancement via Chosen-Plaintext Attack (CPA)}
A critical challenge is determining the optimal decision threshold $\tau$. 
A static heuristic (e.g., $3\sigma$-rule) is often suboptimal due to the varying entropy of different tokens.
However, an adversary with CPA capabilities—who can feed known inputs to the model and observe the resulting KV-cache—can profile the precise statistical behavior of $d_{\text{target}}$ and $d_{\text{other}}$.

This prior knowledge allows for an \textit{adaptive thresholding strategy}. 
We formulate the success probability for a candidate at rank $r$ as:
\begin{equation}
\label{eq:collision_success_rate}
P(\text{success}|r) = P(d_{\text{other}} > \tau)^{r-1} \times P(d_{\text{target}} < \tau).
\end{equation}
This formulation explicitly captures the trade-off: a stricter threshold minimizes False Positives (accepting an incorrect predecessor) but risks False Negatives (rejecting the target). 
By dynamically adjusting $\tau$ based on the expected rank $r$ and the profiled distributions from CPA, the adversary maximizes $P(\text{success})$, significantly boosting reconstruction fidelity compared to static baselines.

\subsubsection{KV-cache Injection Attack}
\label{attack:injection}
Unlike Inversion and Collision attacks which rely on precise algebraic state matching, the \textit{KV-cache Injection Attack} exploits the semantic instruction-following capabilities of LLMs to exfiltrate private contexts.

\partitle{Architectural Constraint: The Missing Query}
Architecturally, a stand-alone KV-cache is computationally dormant. 
The self-attention mechanism requires a current Query vector ($Q$) to drive the attention scoring ($\text{Softmax}(Q K^{\top})$) and state transitions. 
Consequently, an adversary cannot simply ``resume'' inference from a stolen cache; a new input stimulus is required to generate the necessary activation state.

\partitle{Attack Mechanism: Contextual Hijacking}
To overcome this, we weaponize the LLM's instruction-following nature. 
As shown in Figure~\ref{fig:injection}, the adversary appends a crafted directive (e.g., ``\textit{Repeat the previous content.}'') to the intercepted KV-cache. 
This injection serves two purposes:
\textbf{(1) Stimulus Generation.} The directive creates the requisite $Q$ vectors to activate the attention mechanism.
\textbf{(2) Forced Exfiltration.} The LLM attends to the stolen $K/V$ pairs as historical context. 
Bound by its alignment training, the model executes the instruction, effectively ``echoing'' or summarizing the latent private information stored in the cache.

\partitle{Strategic Advantages: Robustness and Efficiency}
This attack offers distinct advantages in robustness compared to the Collision Attack. 
Specifically, it remains effective against KV-cache eviction strategies like H2O~\cite{zhang2023h2o}. 
While such lossy compression breaks the strict mathematical correspondence required for algebraic attacks, it typically preserves the \textit{semantic gist} of the context. 
The Injection Attack successfully exploits these residual semantic vectors to hallucinate or reconstruct private data.
Furthermore, the attack is highly efficient, requiring only a single generation pass. 
This attack underscores that a comprehensive defense must render the KV-cache semantically unintelligible, not just algebraically obfuscated.

\section{Evaluation of Attacks}
\label{sec:experiment}

\subsection{Experimental Setup}
\label{sec:exp_setup}

\partitle{Models} 
To demonstrate attack universality, we select seven state-of-the-art LLMs spanning diverse parameter scales (1B--8B) and attention mechanisms. 
Our evaluation prioritizes modern \textit{GQA} architectures, including the LLaMA-3.2 series (1B \& 3B-Instruct), LLaMA-3.1-8B~\cite{grattafiori2024llama}, and Qwen2.5-Math-7B~\cite{yang2024qwen2}. 
Crucially, we incorporate DeepSeek-R1-Distill-LLaMA-8B~\cite{guo2025deepseek} (LLaMA-3.1-8B-Distilled) to assess robustness against \textit{fine-tuned} models where weight parameters diverge from their base counterparts. 
Finally, to verify generalizability across architectural evolutions, we evaluate legacy \textit{MHA} models, specifically LLaMA-7B and LLaMA-2-7B.

\partitle{Datasets} 
To simulate realistic privacy leakage scenarios, we utilize the LMSYS-Chat-1M dataset~\cite{zheng2023lmsyschat1m}, which comprises real-world user interactions collected from inference services. 
We construct a test set by randomly sampling 1,000 instances, ensuring a rich variety of dialogue and instruction-following contexts. 
Analysis of attack generalization across domain-specific datasets is provided in Appendix~\ref{app:different_datasets}.

\partitle{Evaluation Metrics} 
BERTScore~\cite{zhang2020bertscoreevaluatingtextgeneration} and ROUGE-L~\cite{lin-2004-rouge} are used to measure the similarity between the original input and the text reconstructed from the KV-cache. 
BERTScore, based on the \texttt{all-mpnet-base-v2} model, is better at capturing semantic similarity, while ROUGE-L primarily reflects lexical-level precision and recall.

\subsection{Attack Effectiveness}
\label{sec:experiment_attacks}

\begin{table*}[t!]
    \centering
    \tabcolsep=2.5mm
    \renewcommand{\arraystretch}{1.1}
    \caption{Reconstruction fidelity (BERTScore and ROUGE-L) of attacks against unprotected KV-cache across different models and layers. ``Collision+'' denotes the attack enhanced with prior knowledge.}
    \label{table:attacks}
    \scalebox{0.8}{
        \begin{tabular}{c|c|c|ccc|ccc|ccc|c}
        \hline
        \hline
            \multirow{2}{*}{Type} & \multirow{2}{*}{Model} & \multirow{2}{*}{Metric} & \multicolumn{3}{c|}{Inversion} & \multicolumn{3}{c|}{Collision} & \multicolumn{3}{c|}{Collision+} & Injection \\ \cline{4-13}
            ~ & ~ & ~ & First & Mid & Last & First & Mid & Last & First & Mid & Last & All \\ \hline
            \multirow{10}{*}{Identical} & \multirow{2}{*}{LLaMA-7B} & BERTScore ($\uparrow$) & \cellcolorrr{1.000} & \cellcolorrr{0.065} & \cellcolorrr{0.092} & \cellcolorrr{0.449} & \cellcolorrr{0.769} & \cellcolorrr{0.611} & \cellcolorrr{1.000} & \cellcolorrr{1.000} & \cellcolorrr{1.000} & \cellcolorrr{0.765} \\ 
            ~ & ~ & ROUGE-L ($\uparrow$) & \cellcolorrr{1.000} & \cellcolorrr{0.036} & \cellcolorrr{0.062} & \cellcolorrr{0.500} & \cellcolorrr{0.562} & \cellcolorrr{0.436} & \cellcolorrr{1.000} & \cellcolorrr{1.000} & \cellcolorrr{1.000} & \cellcolorrr{0.687} \\ \cline{2-13}
            ~ & \multirow{2}{*}{LLaMA-3.2-1B} & BERTScore ($\uparrow$) & \cellcolorrr{1.000} & \cellcolorrr{0.084} & \cellcolorrr{0.057} & \cellcolorrr{0.877} & \cellcolorrr{0.791} & \cellcolorrr{0.894} & \cellcolorrr{1.000} & \cellcolorrr{1.000} & \cellcolorrr{1.000} & \cellcolorrr{0.544} \\ 
            ~ & ~ & ROUGE-L ($\uparrow$) & \cellcolorrr{0.994} & \cellcolorrr{0.038} & \cellcolorrr{0.000} & \cellcolorrr{0.709} & \cellcolorrr{0.617} & \cellcolorrr{0.680} & \cellcolorrr{0.994} & \cellcolorrr{0.994} & \cellcolorrr{0.994} & \cellcolorrr{0.315} \\ \cline{2-13}
            ~ & \multirow{2}{*}{LLaMA-3.2-3B-Instruct} & BERTScore ($\uparrow$) & \cellcolorrr{0.055} & \cellcolorrr{0.095} & \cellcolorrr{0.083} & \cellcolorrr{0.782} & \cellcolorrr{0.668} & \cellcolorrr{0.820} & \cellcolorrr{1.000} & \cellcolorrr{1.000} & \cellcolorrr{1.000} & \cellcolorrr{0.540} \\ 
            ~ & ~ & ROUGE-L ($\uparrow$) & \cellcolorrr{0.000} & \cellcolorrr{0.000} & \cellcolorrr{0.000} & \cellcolorrr{0.732} & \cellcolorrr{0.456} & \cellcolorrr{0.621} & \cellcolorrr{0.994} & \cellcolorrr{0.994} & \cellcolorrr{0.994} & \cellcolorrr{0.324} \\ \cline{2-13}
            ~ & \multirow{2}{*}{LLaMA-3.1-8B} & BERTScore ($\uparrow$) & \cellcolorrr{0.071} & \cellcolorrr{0.061} & \cellcolorrr{0.062} & \cellcolorrr{0.873} & \cellcolorrr{0.652} & \cellcolorrr{0.764} & \cellcolorrr{1.000} & \cellcolorrr{1.000} & \cellcolorrr{1.000} & \cellcolorrr{0.616} \\ 
            ~ & ~ & ROUGE-L ($\uparrow$) & \cellcolorrr{0.000} & \cellcolorrr{0.000} & \cellcolorrr{0.001} & \cellcolorrr{0.825} & \cellcolorrr{0.443} & \cellcolorrr{0.564} & \cellcolorrr{0.994} & \cellcolorrr{0.994} & \cellcolorrr{0.994} & \cellcolorrr{0.447} \\ \cline{2-13}
            ~ & \multirow{2}{*}{Qwen2.5-Math-7B} & BERTScore ($\uparrow$) & \cellcolorrr{0.229} & \cellcolorrr{0.105} & \cellcolorrr{0.105} & \cellcolorrr{0.918} & \cellcolorrr{0.552} & \cellcolorrr{0.783} & \cellcolorrr{1.000} & \cellcolorrr{0.983} & \cellcolorrr{0.996} & \cellcolorrr{0.422} \\ 
            ~ & ~ & ROUGE-L ($\uparrow$) & \cellcolorrr{0.186} & \cellcolorrr{0.000} & \cellcolorrr{0.000} & \cellcolorrr{0.842} & \cellcolorrr{0.355} & \cellcolorrr{0.580} & \cellcolorrr{1.000} & \cellcolorrr{0.977} & \cellcolorrr{0.996} & \cellcolorrr{0.286} \\ \hline
            \multirow{2}{*}{Finetune} & \multirow{1}{*}{LLaMA-3.1-8B-Distilled} & BERTScore ($\uparrow$) & \cellcolorrr{0.083} & \cellcolorrr{0.062} & \cellcolorrr{0.081} & \cellcolorrr{0.642} & \cellcolorrr{0.492} & \cellcolorrr{0.635} & \cellcolorrr{0.894} & \cellcolorrr{0.258} & \cellcolorrr{0.762} & \cellcolorrr{0.610} \\ 
            ~ & (LLaMA-3.1-8B) & ROUGE-L ($\uparrow$) & \cellcolorrr{0.000} & \cellcolorrr{0.000} & \cellcolorrr{0.001} & \cellcolorrr{0.633} & \cellcolorrr{0.227} & \cellcolorrr{0.413} & \cellcolorrr{0.868} & \cellcolorrr{0.122} & \cellcolorrr{0.479} & \cellcolorrr{0.421} \\ \hline
            \multirow{2}{*}{Cross-Arch} & \multirow{1}{*}{LLaMA-2-7B} & BERTScore ($\uparrow$) & \cellcolorrr{0.067} & \cellcolorrr{0.086} & \cellcolorrr{0.084} & \cellcolorrr{0.058} & \cellcolorrr{0.070} & \cellcolorrr{0.069} & \cellcolorrr{0.075} & \cellcolorrr{0.062} & \cellcolorrr{0.060} & \cellcolorrr{0.087} \\ 
            ~ & (LLaMA-7B) & ROUGE-L ($\uparrow$) & \cellcolorrr{0.038} & \cellcolorrr{0.057} & \cellcolorrr{0.031} & \cellcolorrr{0.013} & \cellcolorrr{0.000} & \cellcolorrr{0.002} & \cellcolorrr{0.052} & \cellcolorrr{0.017} & \cellcolorrr{0.024} & \cellcolorrr{0.019} \\ \hline
        \hline
        \end{tabular}
    }
    \vspace{-1.2em}
\end{table*}

This section evaluates the feasibility of reconstructing user inputs via the proposed Inversion, Collision, and Injection attacks. 
We systematically test these vectors against the models defined in Section~\ref{sec:exp_setup}. 
Detailed ablation studies are provided in Appendix~\ref{sec:attack_ablation}.

\subsubsection{Experimental Settings} 
Beyond standard reconstruction, we introduce two specific scenarios to evaluate attack robustness under restricted adversarial knowledge:
\textbf{(1)} \textit{Fine-tuning Mismatch:} We attack the fine-tuned \textit{LLaMA-3.1-8B-Distilled} assuming the adversary only possesses weights from its base model (\textit{LLaMA-3.1-8B}). 
This validates effectiveness when exact model weights are unavailable.
\textbf{(2)} \textit{Cross-Architecture Mismatch:} We employ \textit{LLaMA-7B} parameters to attack \textit{LLaMA-2-7B} to assess generalizability across disjoint architectures.

Regarding data scope, the Inversion and Collision attacks operate on single-layer KV-caches. 
To characterize layer-wise vulnerability, we evaluate these attacks on the \textit{first}, \textit{middle}, and \textit{last} layers respectively. 
In contrast, the Injection Attack utilizes the complete, multi-layer KV-cache to leverage the model's full semantic processing capabilities.

\subsubsection{Results of KV-cache Inversion Attack}
As detailed in Table~\ref{table:attacks}, the efficacy of the Inversion Attack generally hinges on two conditions: \textbf{(1)} access to the \textit{first-layer} KV-cache, and \textbf{(2)} an algebraically invertible projection matrix, typical of \textit{MHA}.
Under these constraints, we achieve near-perfect reconstruction fidelity ($\approx 100\%$), whereas deeper layers yield unintelligible noise ($<10\%$) due to semantic fusion.
Notably, the GQA-based LLaMA-3.2-1B also exhibits high first-layer vulnerability ($\approx 100\%$).
We attribute this to the model's high projection rank relative to its hidden state dimension. 
This distinct property allows the \textit{least squares method} to effectively recover inputs despite the non-square nature of the projection matrix.

\begin{takeawaybox}
\textbf{Takeaway 1:} 
The Inversion Attack is effective against the first-layer KV-cache of MHA models, with exceptions for architectures possessing high-rank projection matrices.
\end{takeawaybox}

\subsubsection{Results of KV-cache Collision Attack}
\begin{figure*}[t]    
    \centering            
    \subfloat[The distributions of LLaMA-3.1-8B model.]
    {
        \label{fig:kvcache_distance_distribution_llama3.1-8b}\includegraphics[width=0.45\textwidth]{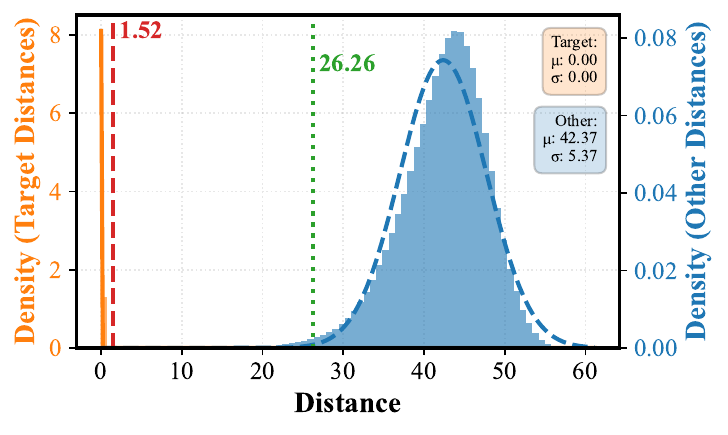}
    }
    \subfloat[The distributions of LLaMA-3.1-8B-Distilled model.]
    {
        \label{fig:kvcache_distance_distribution_dpsk-llama}\includegraphics[width=0.45\textwidth]{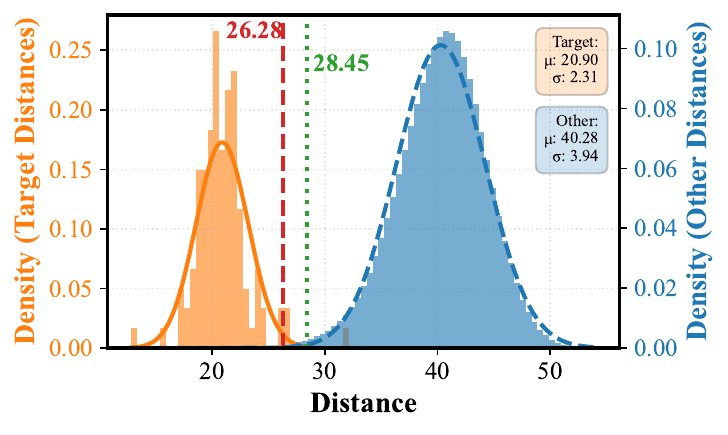}
    }
    \caption{Distance distributions of target tokens $d_{\text{target}}$ (orange) versus incorrect tokens $d_{\text{other}}$ (blue) in the Collision Attack. 
    The input is an excerpt from ``The Bitter Lesson''(see Appendix~\ref{app:bitter_lesson}). 
    The attack targets the last-layer KV-cache of LLaMA-3.1-8B (left) and LLaMA-3.1-8B-Distilled (right) using weights from the base model.
    Vertical lines indicate the heuristic threshold ($3\sigma_{\text{other}}$, green dotted) and the prior-knowledge enhanced threshold ($r=64$, red dashed).}
    \label{fig:kvcache_distance_distribution}
    \vspace{-1.2em}
\end{figure*}

\begin{figure}[t]
    \centering
    \includegraphics[width=0.90\linewidth]{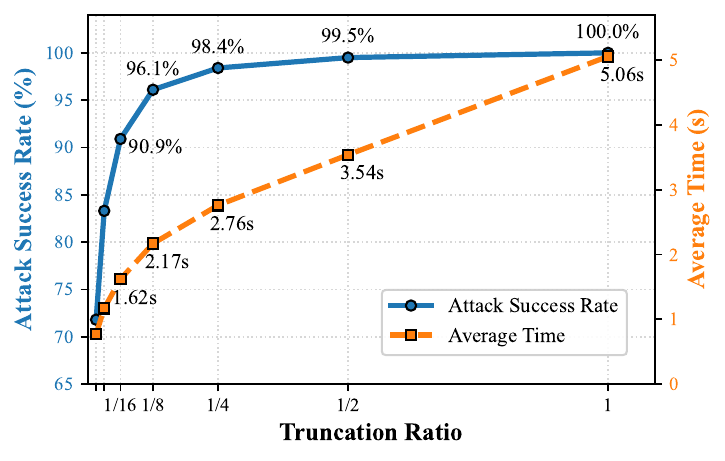}
    \caption{The effect of truncating the probability-sorted vocabulary on reconstruction fidelity and attack time. Experiments were run with a batch size of 256 and an outlier threshold of $3\sigma_{\text{other}}$.}
    \label{fig:stop}
    \vspace{-1.2em}
\end{figure}

Unlike the Inversion Attack, the Collision Attack relies on forward-pass matching using the Frobenius norm.
As shown in Figure~\ref{fig:kvcache_distance_distribution}, we observe that distances for incorrect tokens ($d_{\text{other}}$) approximate a Gaussian distribution, whereas the target token ($d_{\text{target}}$) manifests as a distinct statistical outlier.
By setting an outlier threshold of $3\sigma$ below the batch mean (batch size=256, see Appendix~\ref{app:collision_ablation}), we achieve high reconstruction accuracy across \textit{all} layers of all tested models (Table~\ref{table:attacks}).
Crucially, this method overcomes the architectural constraints of MHA; it is effective even against the fine-tuned \textit{LLaMA-3.1-8B-Distilled} using only public base model weights, demonstrating robustness against parameter divergence.

\partitle{Efficiency via Probability-Guided Pruning}
To enhance practicality, we evaluate truncating the search space based on model-predicted probabilities.
Experimental results on LLaMA-3.2-1B (Figure~\ref{fig:stop}) indicate that searching only the top $1/8$ of tokens retains 96.1\% of the full-search fidelity.
This optimization reduces the average reconstruction time per layer from 5.06s to 2.17s ($<43\%$ of the original latency), rendering the attack highly efficient for real-time scenarios.

\begin{takeawaybox}
\textbf{Takeaway 2:} 
The Collision Attack is a universal threat, achieving high fidelity across diverse architectures (including fine-tuned models) and layers, with high efficiency via probability pruning.
\end{takeawaybox}

\subsubsection{Collision Attack Enhanced with Prior Knowledge (Collision+)}
\label{experiment:collision_enhance}
We further evaluate the potency of integrating adversarial prior knowledge (i.e., adaptive thresholding based on assumed token rank $r$).

\begin{figure}[t]
    \centering
    \includegraphics[width=0.90\linewidth]{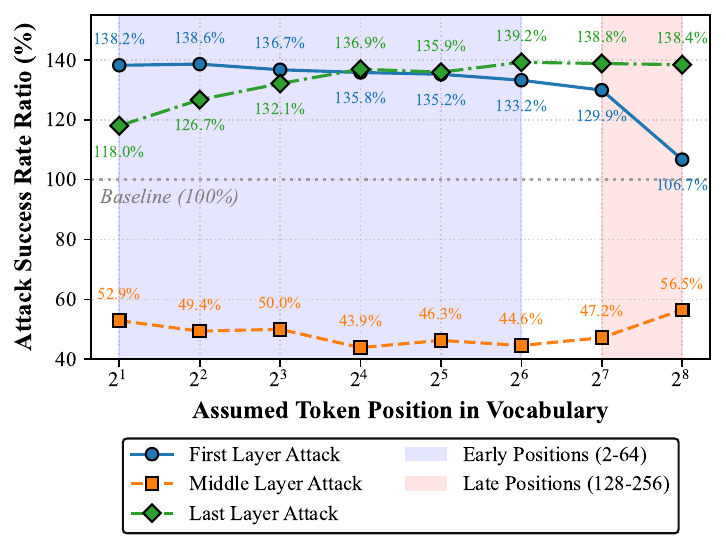}
    \caption{Collision Attack experiments on LLaMA-3.1-8B-Distilled using optimal thresholds derived for different assumed token ranks ($r$).}
    \label{fig:enhance_rank}
    \vspace{-1.2em}
\end{figure}

\partitle{Efficacy on Fine-tuned Models}
As detailed in Table~\ref{table:attacks}, assuming a rank $r=8$ yields near-perfect reconstruction ($\approx 100\%$) across all open-source models. 
For the fine-tuned \textit{LLaMA-3.1-8B-Distilled}, we analyze the impact of rank assumptions in Figure~\ref{fig:enhance_rank}. 
In the \textit{first layer}, where the actual average target rank is $\approx 4$, using the optimal threshold boosts accuracy to 138.6\% of the baseline. 
Conversely, the \textit{middle layer} shows no improvement, implying an actual rank $>128$. 
In the \textit{last layer}, optimal rank alignment ($64 \le r \le 128$) increases accuracy to 139.0\% relative to the baseline.

\partitle{Statistical Justification}
To further illustrate this, we analyze distance distributions using ``The Bitter Lesson'' excerpt (Figure~\ref{fig:kvcache_distance_distribution}). 
For the open-source model, the heuristic $3\sigma_{\text{other}}$ threshold yields a 0.13\% false-positive rate, capping per-token success at 91.84\% (assuming $r=64$). 
By switching to a statistically derived threshold, we eliminate false positives, achieving 100\% success. 
Similarly, for the fine-tuned model, the enhanced threshold improves per-token success from 91.79\% to 97.82\%, dramatically increasing full-sequence reconstruction fidelity.

\begin{takeawaybox}
\textbf{Takeaway 3:} 
Augmenting the Collision Attack with model-specific prior knowledge achieves near-perfect ($\approx100\%$) user-input recovery accuracy.
\end{takeawaybox}

\subsubsection{Results of KV-cache Injection Attack}
We evaluate semantic exfiltration using the optimal directive identified in Appendix~\ref{app:injection_ablation}: ``\textit{Repeat the previous content}''.
As detailed in Table~\ref{table:attacks}, this vector achieves an average BERTScore of 0.58 and ROUGE-L of 0.42.
While these metrics are lower than those of the exact-match Collision Attack, they confirm significant leakage of the input's core meaning.

Notably, the attack exhibits peak performance on \textit{LLaMA-7B}.
This supports the hypothesis that the \textit{MHA} mechanism retains higher contextual fidelity within the KV-cache compared to compressed variants like \textit{GQA}.
Consequently, MHA caches are more semantically intelligible to the model, increasing vulnerability to instruction-based extraction.
This imposes a critical requirement for defenses: a protected KV-cache must be rendered semantically unparsable to the LLM itself, preventing the model from being weaponized to interpret stolen contexts.

\begin{takeawaybox}
\textbf{Takeaway 4:} 
Even without verbatim recovery, Injection Attacks successfully exfiltrate core user intent, necessitating defenses that neutralize semantic intelligibility.
\end{takeawaybox}

\subsection{Attack Robustness under Partial Knowledge Scenarios}

To assess practical applicability, we evaluate the robustness of our attacks under a relaxed threat model characterized by two realistic constraints: incomplete data interception and model parameter mismatch.

\partitle{Incomplete KV-cache Data}
Our analysis reveals distinct data requirements across attack vectors. 
The \textit{Injection Attack} necessitates the full, cross-layer context and fails when data is fragmented.
Conversely, the \textit{Collision Attack} exhibits superior robustness; possessing the KV-cache from \textit{any single layer} is sufficient to perform high-fidelity input reconstruction.

\partitle{Model Mismatch}
We further evaluate the scenario where the adversary's local model differs from the target service model. 
Experiments confirm that attack effectiveness hinges on the correlation of weight parameters.
\begin{itemize}[leftmargin=*]
    \item \textbf{Fine-tuning Mismatch (Gray-box):} We targeted a fine-tuned model (LLaMA-3.1-8B-Distilled) assuming the adversary only possesses the open-source base model (LLaMA-3.1-8B).
    Results in Table~\ref{table:attacks} show that the attack remains highly effective, suggesting that fine-tuning preserves the fundamental weight correlations exploited by our mechanism.
    \item \textbf{Architectural Mismatch (Black-box):} In contrast, cross-architecture attacks (e.g., using LLaMA-7B to attack LLaMA-2-7B) fail completely.
    Reconstruction similarity drops to levels statistically indistinguishable from random guessing ($< 0.1$).
    This confirms that the Collision Attack relies on the algebraic correspondence of weight parameters, which is absent across disjoint architectures.
\end{itemize}

\section{KV-Cloak: A Lightweight KV-cache Defense}
\subsection{Motivation for KV-Cloak}
\label{defense:kv_shield_flaw}

\partitle{Limitations of Existing Privacy-Preserving Techniques} 
Current privacy-preserving techniques fundamentally struggle to balance security, efficiency, and utility when applied to the massive, latency-sensitive KV-cache. We analyze their specific limitations below:
\begin{itemize}[leftmargin=*]
    \item \textbf{Cryptographic Methods (Overhead Bottleneck):} 
    Standard cryptographic techniques~\cite{alenezi2020symmetric, thambiraja2012survey}, such as symmetric AES encryption, provide confidentiality by ensuring data remains in ciphertext during storage and transmission, requiring decryption only for legitimate use. 
    A more advanced approach, Homomorphic Encryption (HE)~\cite{paillier1999public, acar2018survey}, allows for direct computation on encrypted data, theoretically enabling parts of the attention mechanism to operate on the KV-cache without ever decrypting it. 
    However, despite their strong security guarantees, the immense computational overhead and latency introduced by these methods are prohibitive. 
    Given that the KV-cache can be tens or hundreds of gigabytes, applying full encryption or HE is unsuitable for the high-throughput, low-latency requirements of LLM inference.

    \item \textbf{Differential Privacy (Utility Trade-off):} 
    Applying Differential Privacy involves injecting calibrated noise into the Key-Value vectors to mask individual data points~\cite{abadi2016deep, li2025delay, yu2019differentially}. 
    However, the sparse and sensitive nature of attention mechanisms creates a severe utility-privacy trade-off.
    To achieve a meaningful level of privacy, the required amount of noise would significantly degrade the LLM's inference accuracy to an unacceptable degree.

    \item \textbf{KV-Shield (Security and Compatibility Flaws):} 
    KV-Shield~\cite{yang2024first} is the only existing lightweight obfuscation scheme for KV-cache.
    It synchronously permutes attention weight rows ($W_q, W_k, W_v$) to shuffle cache elements, obfuscating them before an attention score is calculated. 
    The obfuscated attention output is then de-obfuscated for subsequent steps. 
    We identify two critical failures in this design: 
    \textbf{(1)} \textit{Security Flaws.} 
    First, the simple shuffling preserves the inherent statistical distribution of the data, leaving it highly vulnerable to our proposed \textit{Collision Attacks}. Second, the reliance on a fixed obfuscation key lacks dynamic randomness, rendering the scheme defenseless against CPA.
    \textbf{(2)} \textit{Architectural Incompatibility.} 
    The shuffling disrupts relative positioning, making it incompatible with RoPE used in SOTA models (e.g., LLaMA, Qwen).
    \label{defense:flaws_of_kvshield}
\end{itemize}

\partitle{Design Objectives} 
The deficiencies of prior works necessitate a specialized defense for KV-cache. 
We define three mandatory design goals for a practical solution:
\begin{itemize}[leftmargin=*]
    \item \textbf{Robust Security:} 
    The mechanism must resist targeted reconstruction attacks by effectively masking both the algebraic and statistical properties of the cache.
    \item \textbf{Lossless Model Fidelity:} 
    It must preserve the exact mathematical equivalence of the attention mechanism, ensuring zero degradation in generation quality.
    \item \textbf{Negligible Overhead:} 
    The defense must operate with minimal latency, ideally shifting computational costs offline to maintain high inference throughput.
\end{itemize}

To meet these objectives, we propose \textbf{KV-Cloak}, a novel defense mechanism that synergizes reversible matrix obfuscation with operator fusion to deliver robust security, lossless model fidelity, and negligible inference overhead.

\subsection{Naive Defense: Reversible Linear Obfuscation}
\label{sec:baseline_defense}

To mitigate privacy risks with minimal overhead, we first consider a naive defense that employs reversible linear transformations to obscure the statistical properties of the KV-cache. 
In the context of modern inference frameworks (e.g., vLLM), we denote the key vectors for a single attention head within a PagedAttention block as a matrix $K \in \mathbb{R}^{b \times d}$, where $b$ is the block size and $d$ is the head dimension. 
Since the protection mechanism applies analogously to value vectors, we focus our analysis on $K$. 
The obfuscation transformation is defined as:
\begin{equation}
    \label{eq:naive_transform}
    K' = SKM,
\end{equation}
where $S \in \mathbb{R}^{b \times b}$ and $M \in \mathbb{R}^{d \times d}$ are secret, randomly generated invertible matrices. 
This transformation aims to conceal the raw content of $K$ while preserving the matrix dimensions necessary for storage allocation.

\partitle{Security Analysis}
Despite its operational simplicity, this scheme is fundamentally insecure under a \textit{CPA} model due to its fixed linear structure.
While a real-world adversary cannot generate arbitrary matrix $K$ directly (as $K$ is constrained by the model's embedding projection of valid tokens), we demonstrate that this constraint does not preclude a full compromise.
An adversary can circumvent this restriction by mounting a \textit{differential attack} through carefully crafted prompts.
Specifically, the adversary chooses two inputs yielding plaintexts $K_1$ and $K_2$, thereby controlling the difference $\Delta K = K_1 - K_2$. 
Due to the linearity of the scheme, the observed ciphertext difference is $\Delta K' = S(\Delta K)M$. 
By systematically crafting inputs such that $\Delta K$ approaches a series of standard basis matrices (i.e., matrices with a single non-zero entry), the adversary can isolate and solve for the columns of $S$ and the rows of $M$. 
This algebraic attack enables full recovery of the secret matrices (up to a scalar ambiguity) with a computational complexity of only $O(b^2d+bd^2)$, rendering the naive defense ineffective against determined adversaries.

\subsection{One-Time Pad Block-wise Shuffling}

\begin{figure}[t]
    \centering
    \includegraphics[width=0.95\linewidth]{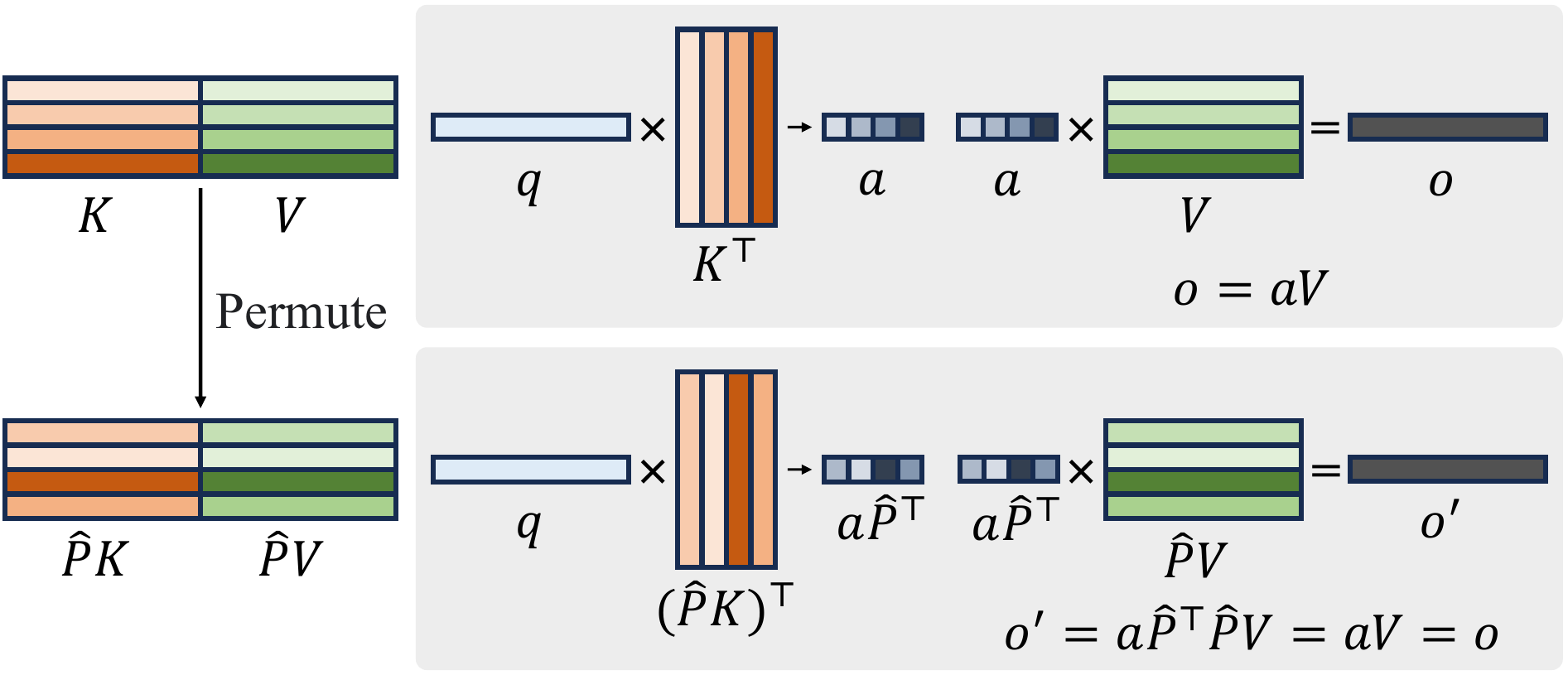}
    \caption{Illustration of the KV-Cloak block-wise shuffling mechanism. By permuting $K$ and $V$ vectors within a block, the mechanism eliminates positional side channels while maintaining mathematical equivalence in the attention output.}
    \label{fig:block_shuffle}
    \vspace{-1.2em}
\end{figure}

\subsubsection{Eliminating Redundant Positional Information in KV-cache}
To enhance the security of the obfuscation scheme, we introduce additional randomness by eliminating the architectural redundancy of the KV-cache. 
We observe that the physical memory ordering of $k, v$ vectors is computationally superfluous during inference, as positional semantics are intrinsic to the vectors via RoPE.

\partitle{Mechanism}
We implement a dynamic, block-wise random permutation. 
By reordering $k, v$ pairs within each block while maintaining their internal correspondence, we decorrelate the physical storage index from the logical token sequence. 
As illustrated in Figure~\ref{fig:block_shuffle}, this operation preserves the mathematical invariance of the attention mechanism while imposing a combinatorial complexity barrier of $b!$ on the adversary. 
Given typical block sizes $b \in \{16, 32, 64\}$~\cite{Kwon2023EfficientMM}, this renders brute-force matching computationally infeasible.

\partitle{Formalization}
The enhanced obfuscation employs a locally generated, one-time pad permutation matrix $\hat{P}$ before applying the linear transformations:
\begin{equation}
    \label{eq:obf_SPM}
    K' = S\hat{P}KM.
\end{equation}
Crucially, for efficiency, $\hat{P}$ is \textit{ephemeral} and does not require storage. 
The inference engine performs subsequent computations directly on the de-obfuscated permuted state $\hat{P}K$ without reconstructing the original order, thereby incurring zero additional storage overhead for the permutation key.

\subsubsection{Rank Preservation and Implicit Key Recovery}  

\begin{figure}[t]
    \centering
    \includegraphics[width=1.0\linewidth]{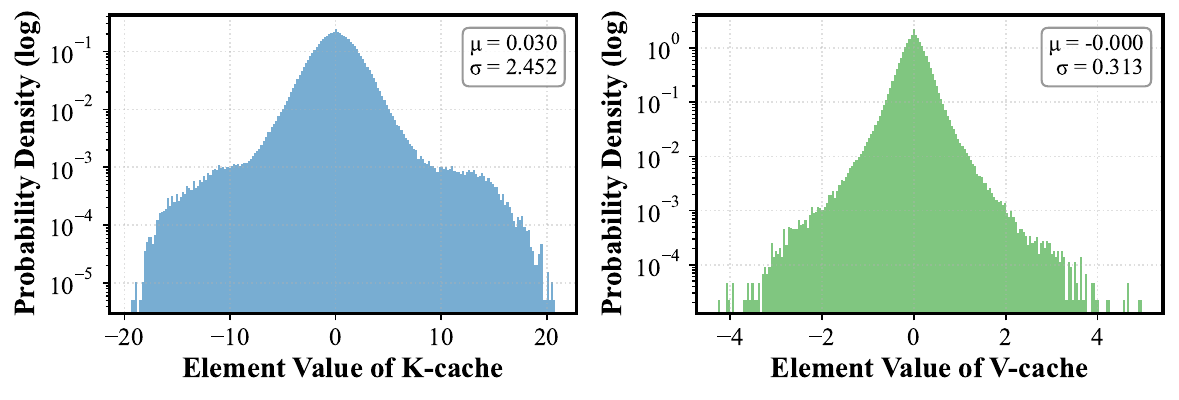}
    \caption{Value distribution of KV-cache elements ($K$ and $V$) for LLaMA-3.2-1B. The data is collected during inference on ``The Bitter Lesson''. 
    }
    \label{fig:kvcache_distribution}
    \vspace{-1.2em}
\end{figure}

A critical challenge arises when the original matrix $K$ exhibits low rank (e.g., rank $1$ in cases of identical repeated tokens). 
In such scenarios, the permutation entropy collapses, as row shuffling becomes statistically invisible. 
To guarantee obfuscation robustness, we introduce a secret additive mask matrix $A$ before permutation.  
Thereby, we can ensure the transformed matrix $(K+A)$ maintains sufficient rank to preserve cryptographic strength of $\hat{P}$.
However, explicitly storing the unique one-time pad $\hat{P}$ for every block would reintroduce significant storage overhead. 
We resolve this by designing $A$ to enable \textit{implicit key recovery} without $\hat{P}$.

\partitle{Magnitude-based Positional Embedding}
We exploit the numerical characteristics of attention activations. 
As shown in Figure~\ref{fig:kvcache_distribution}, elements of $k, v$ vectors are statistically bounded by a threshold $\theta_K$ (e.g., typically $<100$). 
Leveraging this, we construct $A$ to function as a ``positional beacon'' by embedding values significantly larger than $\theta_K$ into specific coordinates. 
This ensures that the matrix $(K+A)$ maintains enough rank.

\partitle{Zero-Storage Recovery}
Since the ``beacons'' in $A$ dominate the magnitude of $K$, the permuted mask term $\hat{P}A$ remains statistically separable from $\hat{P}K$ within the ciphertext $\hat{P}(K+A)$. 
During de-obfuscation, the system dynamically identifies the row permutation indices by detecting these high-magnitude outliers, allowing it to reconstruct $\hat{P}A$ on-the-fly. 
Subtracting this reconstructed mask yields the permuted state $\hat{P}K$ required for inference. 
This mechanism effectively offloads the storage of $\hat{P}$ into the data itself.

\partitle{Formalization}
The complete obfuscation logic integrates this additive masking with the linear and permutation layers:
\begin{equation}
    \label{eq:obf_SPAM}
    K'=S\hat{P}(K+A)M,
\end{equation}
where $\hat{P} \in \{0,1\}^{b \times b}$ is the ephemeral one-time pad, and $A$ is the structured mask. 
This composite transformation simultaneously secures cache statistics and enables efficient, stateless recovery without compromising model accuracy.

\subsubsection{Security Analysis}
The security of KV-Cloak relies fundamentally on the confidentiality of its secret matrices ($S, M, A$). 
Under a standard CPA model, an adversary attempting to recover these keys faces a prohibitive computational barrier. 
Our analysis indicates that the time complexity to solve for the keys is $O((b^2d+bd^2) \cdot b!)$. 
Crucially, the factorial term $b!$, introduced by the dynamic block-wise permutation, renders brute-force key recovery computationally infeasible.

This computational hardness is complemented by layered security mechanisms designed to mitigate specific attack vectors:
\textbf{(1)} The one-time pad permutation severs the explicit token-to-vector positional correspondence at an information-theoretic level, invalidating the premise of Collision Attacks. 
\textbf{(2)} A reversible algebraic transformation via secret matrices ($S, M$) completely disrupts the cache's statistical distribution properties (e.g., mean, variance), neutralizing profiling and data mining attacks. 
\textbf{(3)} The combined obfuscation renders the cache semantically unintelligible and unparsable by pristine (unmodified) models, thereby thwarting Injection Attacks that attempt to leverage the model's own capabilities.

\subsection{Implicit Obfuscation via Operator Fusion}
\label{sec:operator_fusion}

To minimize the runtime latency overhead introduced by the obfuscation and de-obfuscation processes, we propose an \textit{operator fusion} strategy.
This approach integrates the obfuscation operators directly into the model's weight matrices offline, thereby rendering the online computational impact negligible while maintaining strict mathematical equivalence.

\partitle{Transformed Vectors and Invariance}
We introduce two secret, invertible matrices, $M_1, M_2 \in \mathbb{R}^{d \times d}$, to transform the query, key, and value vectors ($q, k, v$) and the output projection $W_o$. 
The transformed vectors ($q^m, k^m, v^m$) and the modified output weight $W_o^m$ are defined as:
\begin{equation}
\label{eq:fused_vectors}
    \left\{\begin{array}{l}
    q^m = q(M_1^{-1})^\top, \\
    k^m = kM_1, \\
    v^m = vM_2, \\
    W_o^m = W_o(M_2^{-1})^\top.
    \end{array}\right.
\end{equation}
This design ensures that the core attention mechanism remains invariant. 
Specifically, the attention scores (scaled dot-product of query and key) are preserved:
\begin{equation}
\label{eq:invariance_attn}
\begin{aligned}
    q_i^m(k_j^m)^\top &= \left(q_i(M_1^{-1})^\top\right) \left(k_jM_1\right)^\top
    = q_i(M_1^{-1})^\top M_1^\top k_j^\top \\
    &= q_i(M_1M_1^{-1})^\top k_j^\top = q_ik_j^\top.
\end{aligned}
\end{equation}
Similarly, the output of the attention head is mathematically identical to the unprotected, guaranteeing lossless accuracy:
\begin{equation}
\label{eq:invariance_output}
\begin{aligned}
    v_j^m(W_o^m)^\top &= (v_jM_2) \left(W_o(M_2^{-1})^\top\right)^\top \\
    &= v_jM_2(M_2^{-1})W_o^\top = v_j W_o^\top.
\end{aligned}
\end{equation}
This invariance guarantees that the model's output is mathematically identical to that of the original, unprotected model, ensuring lossless accuracy.

\partitle{Fusion under RoPE}
Implementing this fusion requires incorporating the input projection $x$ and the position-dependent RoPE matrix $R_{\Theta, i}^d$. The transformed vectors are expressed as:
\begin{equation}
\label{eq:fused_projection}
    \left\{\begin{array}{l}
    q^m = x W_q^\top R_{\Theta, i}^d (M_1^{-1})^\top \\
    k^m = x W_k^\top R_{\Theta, i}^d M_1 \\
    v^m = x W_v^\top M_2 \\
    W_o^m = W_o(M_2^{-1})^\top
    \end{array}\right..
\end{equation}
From Eq.~\ref{eq:fused_projection}, fusing $M_2$ into the value and output weights is straightforward: we pre-compute $(W_v^m)^\top = W_v^\top M_2$ and $W_o^m = W_o(M_2^{-1})^\top$.
However, fusing $M_1$ into $W_k$ and $W_q$ presents a challenge due to the intermediate application of RoPE. 
The fusion is algebraically feasible \textit{if and only if} the secret matrix $M_1$ and the RoPE matrix $R_{\Theta, i}^d$ \textbf{commute}, i.e., $R_{\Theta, i}^d M_1 = M_1 R_{\Theta, i}^d$. 
In this case, Eq.~\ref{eq:fused_projection} simplifies to:
\begin{equation}
    k^m = x W_k^\top (M_1 R_{\Theta, i}^d) = x (W_k^\top M_1) R_{\Theta, i}^d.
\end{equation}
This reordering allows us to absorb $M_1$ into the weight matrix as $(W_k^m)^\top = W_k^\top M_1$. 
A symmetric logic applies to $W_q$.

To satisfy this commutativity constraint, we structurally design $M_1$ as a invertible block-diagonal matrix composed of $2\times2$ rotation-scaling sub-matrices, analogous to the structure of RoPE (detailed in Appendix~\ref{app:commute_with_RoPE}). 
Additionally, we constrain $M_2$ to be a random invertible rotation-scaling matrix and calibrate the scaling factors of both matrices. 
This ensures that after the transformation with mask $A$, the padding tokens remain statistically distinguishable as outliers for implicit key recovery.

\partitle{Offline Computation and Online Efficiency}
Consequently, the obfuscation operators are fully fused into the attention layer parameters offline. 
The new pre-computed weights are:
\begin{equation}
    \label{eq:our_linear_transform}
    \left\{
        \begin{aligned}
            W_q^m &= M_1^{-1} W_q \\
            W_k^m &= M_1^\top W_k \\
            W_v^m &= M_2^\top W_v \\
            W_o^m &= W_o (M_2^{-1})^\top
        \end{aligned}
    \right..
\end{equation}
During online inference, the KV-Cloak obfuscation is applied to the cache generated by these fused weights:
\begin{equation}
    \label{eq:obf_SPA}
    K' = S\hat{P}(K^m+A),
\end{equation}
By eliminating explicit online matrix multiplications with $M$, the cost for protecting one KV-cache block is reduced to the multiplications with $S$, $\hat{P}$, and $S^{-1}$. 
This totals approximately $b^3 + 2b^2d$ floating-point operations. 
For an LLaMA-3.1-8B instance ($b=16, d=128, D=4096$), this overhead represents merely \textbf{0.83\%} of the KV-cache re-computation cost (detailed in Appendix~\ref{sec:performance_analysis}), verifying that KV-Cloak achieves robust security with minimal performance impact.

\begin{table*}[t!]
    \centering
    \tabcolsep=2.5mm
    \renewcommand{\arraystretch}{1.05}
    \caption{Comparison of defense mechanisms (KV-Cloak vs. DP) against input reconstruction attacks on LLaMA-7B, LLaMA-3.2-1B, and LLaMA-3.1-8B-Distilled.}
    \label{table:defense_evaluation}
    \scalebox{0.84}{
        \begin{tabular}{c|c|c|c|ccc|ccc|c}
        \hline
        \hline
            \multirow{2}{*}{Model} & \multirow{2}{*}{Protect Type} & \multirow{2}{*}{Metric} & Inversion & \multicolumn{3}{c|}{Collision} & \multicolumn{3}{c|}{Collision+} & Injection \\ \cline{4-11}
            ~ & ~ & ~ & First & First & Mid & Last & First & Mid & Last & All \\ \hline
            \multirow{6}{*}{LLaMA-7B} & \multirow{2}{*}{Original} & BERTScore ($\downarrow$)& \cellcolorrr{1.000} & \cellcolorrr{0.449} & \cellcolorrr{0.769} & \cellcolorrr{0.611} & \cellcolorrr{1.000} & \cellcolorrr{1.000} & \cellcolorrr{1.000} & \cellcolorrr{0.765} \\ 
            ~ & ~ & ROUGE-L ($\downarrow$)& \cellcolorrr{1.000} & \cellcolorrr{0.500} & \cellcolorrr{0.562} & \cellcolorrr{0.436} & \cellcolorrr{1.000} & \cellcolorrr{1.000} & \cellcolorrr{1.000} & \cellcolorrr{0.687} \\ \cline{2-11}
            ~ & \multirow{2}{*}{KV-Cloak} & BERTScore ($\downarrow$)& \cellcolorrr{0.091} & \cellcolorrr{0.070} & \cellcolorrr{0.069} & \cellcolorrr{0.071} & \cellcolorrr{0.036} & \cellcolorrr{0.036} & \cellcolorrr{0.036} & \cellcolorrr{0.082} \\ 
            ~ & ~ & ROUGE-L ($\downarrow$)& \cellcolorrr{0.068} & \cellcolorrr{0.000} & \cellcolorrr{0.000} & \cellcolorrr{0.000} & \cellcolorrr{0.044} & \cellcolorrr{0.044} & \cellcolorrr{0.044} & \cellcolorrr{0.000} \\ \cline{2-11}
            ~ & \multirow{2}{*}{($10^{8},10^{-5}$)-DP} & BERTScore ($\downarrow$)& \cellcolorrr{0.085} & \cellcolorrr{0.082} & \cellcolorrr{0.672} & \cellcolorrr{0.344} & \cellcolorrr{0.109} & \cellcolorrr{0.937} & \cellcolorrr{0.991} & \cellcolorrr{0.085} \\ 
            ~ & ~ & ROUGE-L ($\downarrow$)& \cellcolorrr{0.065} & \cellcolorrr{0.041} & \cellcolorrr{0.433} & \cellcolorrr{0.197} & \cellcolorrr{0.097} & \cellcolorrr{0.901} & \cellcolorrr{0.979} & \cellcolorrr{0.009} \\ \hline
            \multirow{6}{*}{LLaMA-3.2-1B} & \multirow{2}{*}{Original} & BERTScore ($\downarrow$)& \cellcolorrr{1.000} & \cellcolorrr{0.877} & \cellcolorrr{0.791} & \cellcolorrr{0.894} & \cellcolorrr{1.000} & \cellcolorrr{1.000} & \cellcolorrr{1.000} & \cellcolorrr{0.544} \\ 
            ~ & ~ & ROUGE-L ($\downarrow$)& \cellcolorrr{0.994} & \cellcolorrr{0.709} & \cellcolorrr{0.617} & \cellcolorrr{0.680} & \cellcolorrr{0.994} & \cellcolorrr{0.994} & \cellcolorrr{0.994} & \cellcolorrr{0.315} \\ \cline{2-11}
            ~ & \multirow{2}{*}{KV-Cloak} & BERTScore ($\downarrow$)& \cellcolorrr{0.085} & \cellcolorrr{0.072} & \cellcolorrr{0.074} & \cellcolorrr{0.069} & \cellcolorrr{0.051} & \cellcolorrr{0.051} & \cellcolorrr{0.051} & \cellcolorrr{0.079} \\ 
            ~ & ~ & ROUGE-L ($\downarrow$)& \cellcolorrr{0.009} & \cellcolorrr{0.000} & \cellcolorrr{0.000} & \cellcolorrr{0.000} & \cellcolorrr{0.002} & \cellcolorrr{0.002} & \cellcolorrr{0.002} & \cellcolorrr{0.000} \\ \cline{2-11}
            ~ & \multirow{2}{*}{($10^{8},10^{-5}$)-DP} & BERTScore ($\downarrow$)& \cellcolorrr{0.633} & \cellcolorrr{0.849} & \cellcolorrr{0.763} & \cellcolorrr{0.849} & \cellcolorrr{0.973} & \cellcolorrr{0.995} & \cellcolorrr{1.000} & \cellcolorrr{0.393} \\ 
            ~ & ~ & ROUGE-L ($\downarrow$)& \cellcolorrr{0.622} & \cellcolorrr{0.604} & \cellcolorrr{0.587} & \cellcolorrr{0.604} & \cellcolorrr{0.966} & \cellcolorrr{0.989} & \cellcolorrr{0.994} & \cellcolorrr{0.248} \\ \hline
            \multirow{6}{*}{LLaMA-3.1-8B-Distilled} & \multirow{2}{*}{Original} & BERTScore ($\downarrow$)& \cellcolorrr{0.083} & \cellcolorrr{0.642} & \cellcolorrr{0.492} & \cellcolorrr{0.635} & \cellcolorrr{0.885} & \cellcolorrr{0.251} & \cellcolorrr{0.829} & \cellcolorrr{0.610} \\ 
            ~ & ~ & ROUGE-L ($\downarrow$)& \cellcolorrr{0.000} & \cellcolorrr{0.633} & \cellcolorrr{0.227} & \cellcolorrr{0.413} & \cellcolorrr{0.858} & \cellcolorrr{0.112} & \cellcolorrr{0.552} & \cellcolorrr{0.421} \\ \cline{2-11}
            ~ & \multirow{2}{*}{KV-Cloak} & BERTScore ($\downarrow$)& \cellcolorrr{0.093} & \cellcolorrr{0.070} & \cellcolorrr{0.070} & \cellcolorrr{0.069} & \cellcolorrr{0.041} & \cellcolorrr{0.041} & \cellcolorrr{0.041} & \cellcolorrr{0.088} \\ 
            ~ & ~ & ROUGE-L ($\downarrow$)& \cellcolorrr{0.002} & \cellcolorrr{0.000} & \cellcolorrr{0.000} & \cellcolorrr{0.000} & \cellcolorrr{0.003} & \cellcolorrr{0.003} & \cellcolorrr{0.003} & \cellcolorrr{0.000} \\ \cline{2-11}
            ~ & \multirow{2}{*}{($10^{8},10^{-5}$)-DP} & BERTScore ($\downarrow$)& \cellcolorrr{0.079} & \cellcolorrr{0.320} & \cellcolorrr{0.440} & \cellcolorrr{0.566} & \cellcolorrr{0.526} & \cellcolorrr{0.267} & \cellcolorrr{0.824} & \cellcolorrr{0.118} \\ 
            ~ & ~ & ROUGE-L ($\downarrow$)& \cellcolorrr{0.003} & \cellcolorrr{0.291} & \cellcolorrr{0.185} & \cellcolorrr{0.351} & \cellcolorrr{0.530} & \cellcolorrr{0.122} & \cellcolorrr{0.543} & \cellcolorrr{0.049} \\ \hline
        \hline
        \end{tabular}
    }
    \vspace{-1.0em}
\end{table*}

\section{Evaluation of KV-Cloak}

\subsection{Experimental Settings}
We perform a systematic evaluation across three dimensions: \textit{Security}, \textit{Model Accuracy}, and \textit{Performance Overhead}. 
To benchmark the efficacy of KV-Cloak, we compare it against three distinct baselines: 
\textbf{(1)} a standard, unprotected system (Plaintext), 
\textbf{(2)} a defense based on differential privacy with Gaussian noise (DP), and \textbf{(3)} a defense via standard cryptographic encryption (AES). 

\partitle{KV-Cloak Configuration}
Our implementation involves matrix multiplications with invertible secret matrices $S$ and $M$, and an additive mask $A$. 
We configure these parameters to balance security and numerical stability:
\begin{itemize}[leftmargin=*]
    \item \textbf{Block Size $b$:} 
    To maintain compatibility with the PagedAttention mechanism (e.g., vLLM), we experimented with standard block sizes of $b \in \{16, 32, 64\}$.
    \item \textbf{Secret Matrices $S,M$:} 
    To minimize precision loss during the matrix inversion required for de-obfuscation, we sample $S$ and $M$ strictly from the orthogonal group.
    \item \textbf{Additive Mask $A$ and Padding:} 
    To preserve numerical precision, the additive mask $A$ and padding values are magnitude-constrained. 
    We sample elements of $A$ uniformly from $[3\theta_K, 4\theta_K]$ and set padding to $1.5\theta_K$. 
    Here, $\theta_K$ denotes the maximum absolute value observed in the K-cache during calibration on an excerpt from ``The Bitter Lesson''. 
    This methodology is symmetrically applied to the V-cache.
\end{itemize}

\partitle{DP Baseline Configuration}
To establish a strong DP baseline, we conducted an ablation study (detailed in Appendix~\ref{app:dp_ablation}) to optimize the trade-off between privacy and utility. 
We set the clipping threshold to the 50th percentile of the L2 norm distribution observed across the dataset. 
Based on this, we apply Gaussian noise calibrated for $(\epsilon=10^{8}, \delta=10^{-5})$-DP independently to the K/V caches. 
We explicitly select $\epsilon=10^8$ because stricter privacy budgets (e.g., $\epsilon=10^7$) resulted in near-zero inference accuracy (see Table~\ref{table:dp_accuracy}).

\subsection{Evaluation of Security}

\begin{figure*}[t]
    \centering
    \subfloat[The distributions of LLaMA-3.2-1B model without protection.]
    {
        \label{fig:kvcache_distance_distribution_llama3.2-1b}\includegraphics[width=0.48\textwidth]{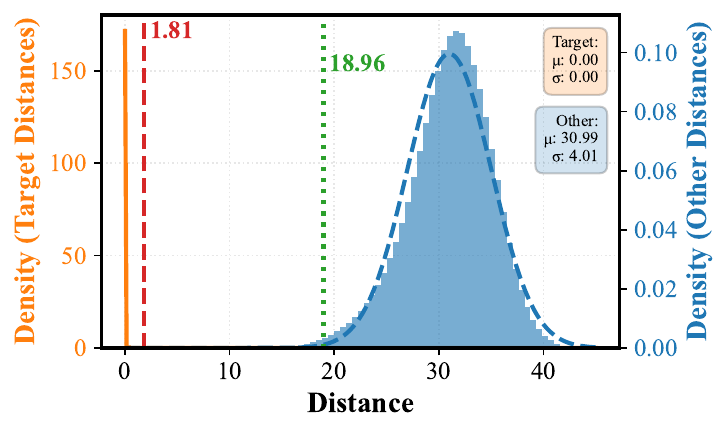}
    }
    \subfloat[The distributions of LLaMA-3.2-1B model with KV-Cloak.]
    {
        \label{fig:kvcache_distance_distribution_llama3.2-1b_kvcloak}\includegraphics[width=0.48\textwidth]{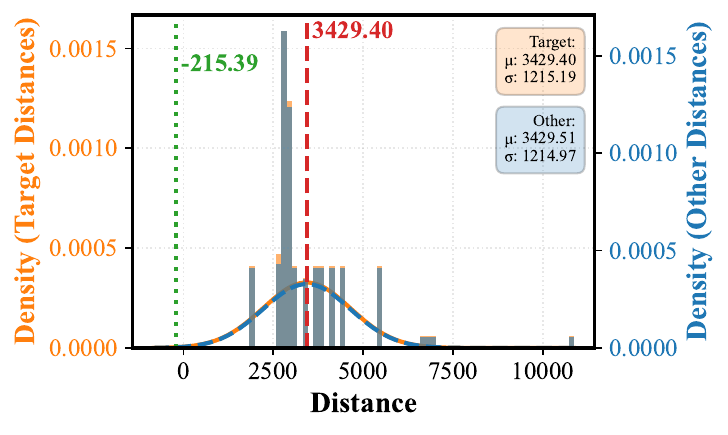}
    } \\
    \subfloat[The distributions of LLaMA-3.2-1B model with ($10^{7},10^{-5}$)-DP.]
    {
        \label{fig:kvcache_distance_distribution_llama3.2-1b_1e7dp}\includegraphics[width=0.48\textwidth]{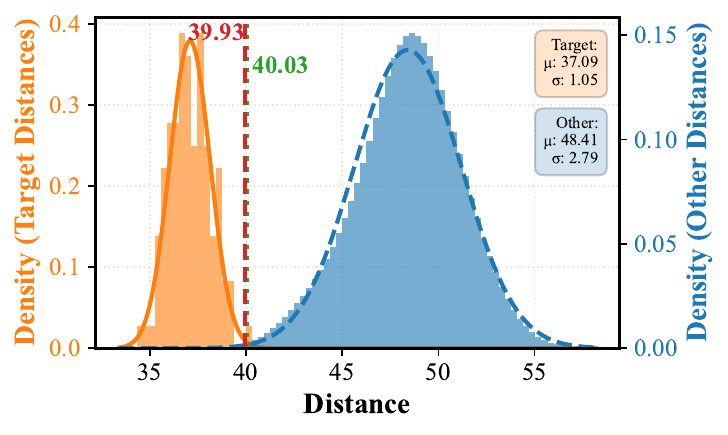}
    }
    \subfloat[The distributions of LLaMA-3.2-1B model with ($10^{8},10^{-5}$)-DP.]
    {
        \label{fig:kvcache_distance_distribution_llama3.2-1b_1e8dp}\includegraphics[width=0.48\textwidth]{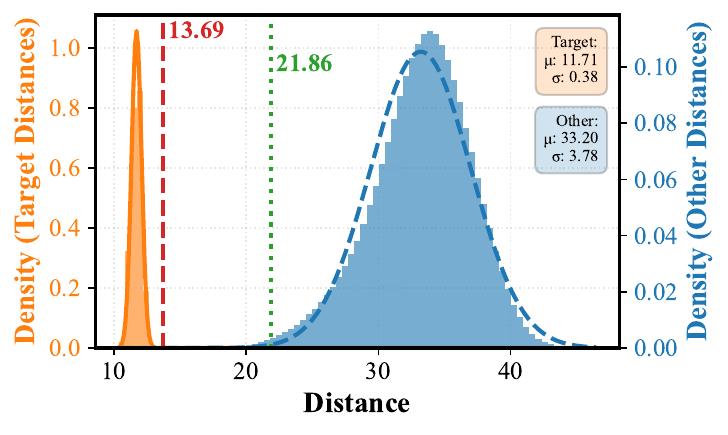}
    }
    \caption{Distance distributions of target tokens $d_{\text{target}}$ (orange) versus incorrect tokens $d_{\text{other}}$ (blue) in the Collision Attack. 
    The experiment targets the last-layer KV-cache of LLaMA-3.2-1B using an excerpt from ``The Bitter Lesson''.
    Subplots display distributions under four conditions: (a) Plaintext, (b) KV-Cloak protection, (c) ($10^7,10^{-5}$)-DP protection, and (d) ($10^8,10^{-5}$)-DP protection. 
    Vertical lines indicate the heuristic threshold ($3\sigma_{\text{other}}$, green dotted) and the prior-knowledge enhanced threshold ($r=64$, red dashed).}
    \label{fig:kvcache_distance_distribution_protected}
    \vspace{-1.2em}
\end{figure*}

This section evaluates the capability of KV-Cloak against the three proposed reconstruction attacks. 
We benchmark against the Plaintext baseline and Differential Privacy mechanisms, applying the attacks to the protected KV-cache.

\partitle{Defense Efficacy Comparison}
As summarized in Table~\ref{table:defense_evaluation} (evaluation on the remaining models can be found in Appendix \ref{app:defense_evaluation_other}), the \textit{Plaintext} baseline is highly vulnerable, yielding high attack success rates on all three attacks. 
In stark contrast, applying KV-Cloak drastically degrades reconstruction quality. 
The BERTScore of all attack outputs drops to a level consistent with random chance, and the ROUGE-L score falls to nearly $0$. 
These results are statistically indistinguishable from comparing the original input with a random string, demonstrating that semantic reconstruction is entirely disrupted. 
This proves that KV-Cloak effectively protects the private information within the KV-cache. 
For the \textit{DP baseline}, security is heavily dependent on the privacy budget $\epsilon$. 
With a weak budget of $\epsilon=10^8$, while Inversion and Injection attacks are mitigated, the Collision Attack remains effective, recovering substantial private information.

\partitle{Statistical Indistinguishability Analysis}
To understand the root cause of defense failure or success, we analyze the distance distributions of target tokens ($d_{\text{target}}$) versus other tokens ($d_{\text{other}}$) for the Collision Attack on LLaMA-3.2-1B (Fig.~\ref{fig:kvcache_distance_distribution_protected}).
Under DP protection, the two distributions remain statistically distinguishable. 
Consequently, utilizing an enhanced threshold with prior knowledge yields a per-token success rate of 94.84\% for ($10^7, 10^{-5}$)-DP, rising to nearly 100\% (equivalent to Plaintext) for ($10^8, 10^{-5}$)-DP.
Conversely, KV-Cloak achieves \textit{distribution collapse}: the distributions of $d_{\text{target}}$ and $d_{\text{other}}$ become completely indistinguishable. 
This eliminates the statistical separability required for the attack, resulting in a $0\%$ success rate.

\partitle{Robustness against Enhanced Attacks}
To assess the robustness of our defense, We further evaluate by subjecting KV-Cloak to the enhanced Collision Attack (utilizing adversarial prior knowledge) across multiple LLMs. 
As shown in Table~\ref{table:defense_evaluation}, KV-Cloak demonstrates consistent resilience. 
Even against this advanced vector, reconstruction accuracy remains near-zero, and outputs are qualitatively equivalent to random noise. 
This confirms that KV-Cloak effectively neutralizes both algebraic and statistical attack vectors.

\begin{takeawaybox}
\textbf{Takeaway 5:} 
KV-Cloak completely thwarts all proposed attacks, reducing the quality of reconstructed text to a level statistically indistinguishable from random noise.
\end{takeawaybox}

\subsection{Inference Accuracy}
\label{sec:experiment_defense_accuracy}

\begin{table}[t]
    \centering
    \tabcolsep=0.6mm
    \renewcommand{\arraystretch}{1.4}
    \caption{Impact of KV-Cloak on inference accuracy (higher is better) across various models, using a block size of 16.}
    \label{table:accuracy}
    \scalebox{0.78}{
        \begin{tabular}{c|cc|cc|cc}
        \hline
        \hline
            \multirow{2}{*}{Model} & \multicolumn{2}{c|}{Plaintext} & \multicolumn{2}{c|}{KV-Cloak} & \multicolumn{2}{c}{($10^8,10^{-5}$)-DP} \\ \cline{2-7}
            ~ & MMLU & SQuAD & MMLU & SQuAD & MMLU & SQuAD \\ \hline
            LLaMA-7B & 0.304 & 0.646 & 0.304 & 0.652 & 0.016 & 0.000 \\ \cline{1-7}
            LLaMA-3.2-1B & 0.335 & 0.457 & 0.335 & 0.458 & 0.262 & 0.258 \\ \cline{1-7}
            LLaMA-3.2-3B-Instruct & 0.619 & 0.652 & 0.619 & 0.652 & 0.379 & 0.012 \\ \cline{1-7}
            LLaMA-3.1-8B & 0.668 & 0.708 & 0.668 & 0.709 & 0.283 & 0.026 \\ \cline{1-7} 
            LLaMA-3.1-8B-Distilled & 0.584 & 0.568 & 0.584 & 0.570 & 0.108 & 0.001 \\ \cline{1-7}
            Qwen2.5-Math-7B & 0.620 & 0.630 & 0.620 & 0.632 & 0.042 & 0.000 \\ \hline
        \hline
        \end{tabular}
    }
    \vspace{-1.2em}
\end{table}

To rigorously assess model fidelity, we simulate a disaggregated inference service employing a prefill-decode architecture. 
In this pipeline, the KV-cache generated during the prefill phase is secured using either DP or KV-Cloak, transferred, and subsequently utilized by the decode node for token generation. 
We employ two standard benchmarks to measure utility: \textit{MMLU}~\cite{hendryckstest2021, hendrycks2021ethics} for massive multitask knowledge and \textit{SQuAD}~\cite{rajpurkar-etal-2016-squad} for reading comprehension.

We evaluated KV-Cloak across all experimental models (results detailed in Table~\ref{table:accuracy}).
The empirical data corroborates our theoretical design: KV-Cloak exhibits \textit{zero degradation} in model performance. 
Unlike Differential Privacy, which forces a trade-off between utility and privacy via noise injection, KV-Cloak leverages reversible linear transformations. 
This ensures that the de-obfuscated attention states are mathematically identical to the plaintext baseline. 
Consequently, the impact on inference accuracy is negligible across both benchmarks, confirming that KV-Cloak is a practically lossless solution.

\begin{takeawaybox}
\textbf{Takeaway 6:} 
KV-Cloak is virtually lossless, preserving the model's fidelity and core utility without any degradation.
\end{takeawaybox}

\subsection{Evaluation of Performance Overhead}
\label{sec:experiment_defense_overhead}

\begin{figure}[t]
    \centering
    \includegraphics[width=0.9\linewidth]{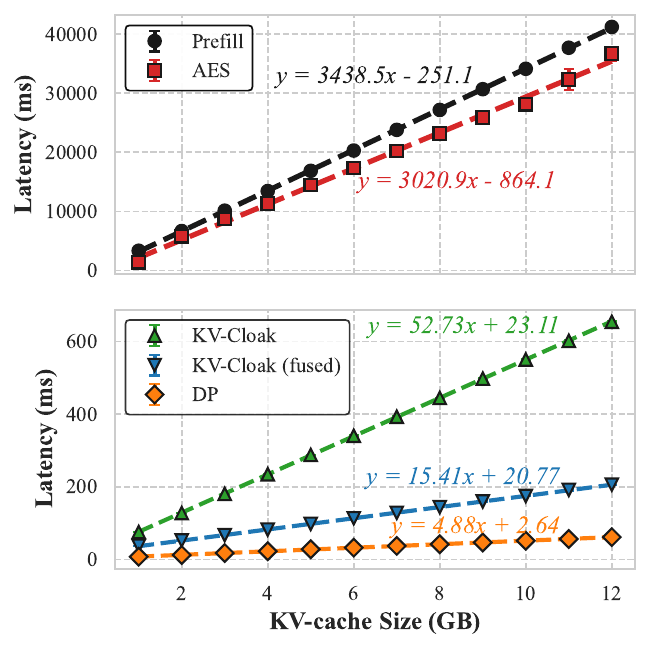}
    \caption{
    Micro-benchmark of latency overhead (ms/GB) versus KV-cache size on LLaMA-3.1-8B. 
    }
    \label{fig:micro_latency}
    \vspace{-1.2em}
\end{figure}

\partitle{Computational Overhead}
To accurately assess the upper bound of performance impact, we employ a conservative, worst-case micro-benchmark simulating a prefill-decode split. 
Measurements are conducted on a serial PyTorch implementation without custom CUDA kernels, isolating the algorithmic overhead.
As illustrated in Figure~\ref{fig:micro_latency}, latency scales linearly. 
Standard AES encryption incurs a prohibitive cost of 3020.9 ms/GB, nearly rivaling the prefill latency (3438.5 ms/GB) itself.
In stark contrast, KV-Cloak with operator fusion introduces a negligible overhead of just \textbf{15.41 ms/GB}. 
This constitutes merely \textbf{0.45\%} of the prefill latency, ranking second only to the insecure DP scheme (4.88 ms/GB). 
In practical deployments, unavoidable network latency would further mask this minimal computation cost, confirming the efficiency of our design.

\partitle{Storage Overhead}
The storage Overhead for KV-Cloak's secret matrices ($S, M, A$) is negligible. 
The total size, $2n_{\text{kv}}l(b^2+\frac{3}{2}d+1)$, is orders of magnitude smaller than the protected cache. 
Quantitatively, for LLaMA-3.1-8B (using block size $b=16$), the overhead is merely 898 KB; even for a 1T-parameter model like Ling-1T~\cite{li2025every} (scaling to $b=64$), it remains only 20.9 MB.
This MB-scale footprint allows all cryptographic keys to securely reside within the limited memory of TEEs with negligible management overhead.

\begin{takeawaybox}
\textbf{Takeaway 7:} KV-Cloak introduces negligible computational and storage overhead to the inference pipeline.
\end{takeawaybox}

\section{Discussion and Future Work}
While KV-Cloak provides a solid foundation for protecting the KV-cache, we identify several limitations, which in turn open up exciting avenues for future research.

\partitle{Key Management and Hardware Security Integration}
The security of KV-Cloak fundamentally relies on the confidentiality of its (megabyte-scale) key matrices. 
Currently, we assume these secrets are protected via TEEs. 
Future work should deeply integrate KV-Cloak with hardware-level security mechanisms, such as TEEs or confidential GPUs, to construct a more robust defense-in-depth architecture.
Furthermore, to counter long-term cryptanalysis in persistent services, developing efficient, low-latency key rotation protocols is essential.

\partitle{Performance Optimization via Co-Design}
Although our evaluation shows minimal overhead, hyper-scale deployment demands further optimization. 
At the software level, latency can be masked through the asynchronous generation of One-Time Pad (OTP) matrices, decoupling security operations from the critical inference path. 
At the hardware level, an algorithm-hardware co-design approach—implementing dedicated GPU intrinsics for block-wise permutation and matrix multiplication—could render the obfuscation cost virtually transparent.

\partitle{Adaptation to Quantized Models}
Our current prototype targets floating-point models. 
As the industry shifts to integer quantization (e.g., INT8 or INT4) for efficiency, extending KV-Cloak becomes a priority. 
This requires designing new lossless reversible transformations for discrete data types, potentially based on structures like modular arithmetic.

\section{Conclusion}
\label{sec:conclusion}

This research exposes a critical security flaw at the heart of modern LLM inference systems: the privacy risk of data leakage from the KV-cache. We have demonstrated the feasibility of reconstructing sensitive user inputs through three novel attack strategies, with our Collision Attack proving particularly effective across various models. This underscores the urgent need for dedicated protection mechanisms that do not compromise the efficiency gains the KV-cache provides.

In response, we designed KV-Cloak. By employing a lightweight, reversible obfuscation technique, KV-Cloak neutralizes the identified threats without degrading model accuracy or imposing significant latency. It is designed for seamless integration into existing high-performance inference frameworks like vLLM. Our work provides a vital contribution to building secure and trustworthy AI, offering a blueprint for balancing the competing demands of performance and user privacy in the next generation of LLM services. It establishes that strong privacy protection can be achieved without sacrificing the utility and efficiency that have made these models so powerful.

\section*{Acknowledgment}

This research is supported in part by the ``Pioneer'' and ``Leading Goose'' R\&D Program of Zhejiang (2024C01169), and the National Natural Science Foundation of China under Grants (62441238, U2441240).
\clearpage

\section*{Ethic Considerations}

\label{sec:ethics}

This research aims to enhance the privacy and trustworthiness of LLM inference, but we acknowledge the dual-use nature of the vulnerabilities and attack methods we have uncovered. To fulfill our ethical responsibilities and mitigate any potential for misuse, we have committed to a policy of responsible disclosure. Prior to the public dissemination of this paper, we will share our findings, including the details of the vulnerabilities and our proposed defense, with the developers of major affected inference frameworks, such as vLLM. Furthermore, all of our attack validation experiments were conducted exclusively on public and non-sensitive academic datasets. No real user data was involved at any stage of our research. We firmly believe that by taking these responsible measures, the positive contributions of our defensive work, KV-Cloak, will far outweigh the risks associated with the disclosure of these attacks. We are confident that this work will encourage the community to build more trustworthy AI services that are secure by default.

\appendix

\subsection{Proof of Commutativity with Rotary Position Embedding}
\label{app:commute_with_RoPE}
Denoting both ${R}_{\Theta, i}^d$ and the random invertible matrix $M_1$ as a $2\times 2$ block matrix, we obtain:
\begin{equation}
    \label{eq:RoPE_block}
    {R}_{\Theta, i}^d =\left[\begin{array}{cc}
    C & -S \\
    S & C
    \end{array}\right],
    M_1 =\left[\begin{array}{cc}
    T & Y \\
    U & Z 
    \end{array}\right],
\end{equation}
where $C,S\in \mathbb{R}^{\frac{d}{2}\times\frac{d}{2}}$ are shown in Eq.~(\ref{eq:RoPE_block_C_S}), assuming that $T,U,Y,Z \in \mathbb{R}^{\frac{d}{2}\times\frac{d}{2}}$ are all random matrices.
\begin{equation}
    \label{eq:RoPE_block_C_S}
    \begin{aligned}
    C &= \left[\begin{array}{cccc}
    \cos i \theta_0 & 0 & \cdots & 0 \\
    0 & \cos i \theta_1 & \cdots & 0 \\
    \vdots & \vdots & \ddots & \vdots \\
    0 & 0 & \cdots & \cos i \theta_{\frac{d}{2}-1} 
    \end{array}\right],\\
    S &= \left[\begin{array}{cccc}
    \sin i \theta_0 & 0 & \cdots & 0 \\
    0 & \sin i \theta_1 & \cdots & 0 \\
    \vdots & \vdots & \ddots & \vdots \\
    0 & 0 & \cdots & \sin i \theta_{\frac{d}{2}-1} 
    \end{array}\right].
    \end{aligned}
\end{equation}
If the secret matrix $M_1$ and the RoPE matrix $R_{\Theta, i}^d$ commute, then:
\begin{equation}
    \label{eq:commutative_property_block}
    \left[\begin{array}{cc}
    T & Y \\
    U & Z 
    \end{array}\right] \left[\begin{array}{cc}
    C & -S \\
    S & C
    \end{array}\right] = \left[\begin{array}{cc}
    C & -S \\
    S & C
    \end{array}\right] \left[\begin{array}{cc}
    T & Y \\
    U & Z 
    \end{array}\right]. 
\end{equation}

With $j$ and $k$ as subscripts of matrix elements, Eq.~(\ref{eq:commutative_property_linear_equations}) is equivalent to the following system of linear equations:
\begin{equation}
    \label{eq:commutative_property_linear_equations}
    \left\{\begin{array}{l}
    t_{j k}\cos i\theta_j - u_{j k} \sin i\theta_j =t_{j k} \cos i\theta_k+y_{j k} \sin i\theta_k \\
    t_{j k}\sin i\theta_j + u_{j k}\cos i\theta_j =u_{j k}\cos i\theta_k+z_{j k}\sin i\theta_k \\
    y_{j k}\cos i\theta_j -z_{j k}\sin i\theta_j =-t_{j k} \sin i\theta_k+y_{j k} \cos i\theta_k \\
    y_{j k}\sin i\theta_j +z_{j k}\cos i\theta_j=-u_{j k} \sin i\theta_k+z_{j k} \cos i\theta_k
    \end{array}\right.
\end{equation}
Calculating the equation, the $T, U, Y, Z$ need to satisfy the following relationship:
\begin{equation}
    \label{eq:commutative_property_solve}
    \left\{\begin{array}{l}
        y_{jj}=-u_{jj},\\
        z_{jj}=t_{jj},\\
        t_{jk}=u_{jk}=y_{jk}=z_{jk}=0, (j\neq k)
    \end{array}\right..
\end{equation}
Specifically, the structure of $M_1$ is defined as follows in Eq.~(\ref{eq:commutative_property_solve_2}),
\begin{equation}
    \label{eq:commutative_property_solve_2}
    M_1 =\left[\begin{array}{cccccccc}
        t_0 & 0 & \cdots & 0 & -u_0 & 0 & \cdots & 0 \\
        0 & t_1 & \cdots & 0 & 0 & -u_1 & \cdots & 0 \\
        \vdots & \vdots & \ddots & \vdots & \vdots & \vdots & \ddots & \vdots \\
        0 & 0 & \cdots & t_{\frac{d}{2}-1} & 0 & 0 & \cdots & -u_{\frac{d}{2}-1} \\
        u_0 & 0 & \cdots & 0 & t_0 & 0 & \cdots & 0 \\
        0 & u_1 & \cdots & 0 & 0 & t_1 & \cdots & 0 \\
        \vdots & \vdots & \ddots & \vdots & \vdots & \vdots & \ddots & \vdots \\
        0 & 0 & \cdots & u_{\frac{d}{2}-1} & 0 & 0 & \cdots & t_{\frac{d}{2}-1}
        \end{array}\right].
\end{equation}

\subsection{Input Text for Parameter Calibration}
\label{app:bitter_lesson}
The following text, an excerpt from ``The Bitter Lesson'' by Rich Sutton, was used as model input in our experiments to analyze the numerical characteristics of the KV-cache (e.g., for parameter calibration as described in Section~\ref{experiment:collision_enhance}).

\begin{tcolorbox}[
    colback=black!5!white,
    colframe=black!40!white,
    fonttitle=\bfseries,
    title=Input Text for Parameter Calibration,
    arc=4mm,
]
    One thing that should be learned from the bitter lesson is the great power of general-purpose methods, of methods that continue to scale with increased computation even as the available computation becomes very great. The two methods that seem to scale arbitrarily in this way are search and learning. The second general point to be learned from the bitter lesson is that the actual contents of minds are tremendously, irredeemably complex; we should stop trying to find simple ways to think about the contents of minds, such as simple ways to think about space, objects, multiple agents, or symmetries.
\end{tcolorbox}

\subsection{Additional Experiments about Attacks}
\label{sec:attack_ablation}

\subsubsection{Ablation Study of Collision Attack}
\label{app:collision_ablation}

\begin{table}[t]
    \centering
    \tabcolsep=1.0mm
    \renewcommand{\arraystretch}{1.1}
    \caption{Impact of different outlier detection thresholds on reconstruction accuracy, with a fixed batch size of 256.}
    \label{table:gap}
    \scalebox{0.79}{
        \begin{tabular}{c|c|c|ccccc}
        \hline
        \hline
            \multirow{2}{*}{Model} & \multirow{2}{*}{Layer} & \multirow{2}{*}{Metric} & \multicolumn{5}{c}{Gap} \\ \cline{4-8}
            ~ & ~ & ~ & $2\sigma$ & $2.5\sigma$ & $3\sigma$ & $3.5\sigma$ & $4\sigma$ \\ \hline
            \multirow{7}{*}{LLaMA-3.2-1B} & \multirow{2}{*}{First} & BERTScore ($\uparrow$) & \cellcolorrr{0.531} & \cellcolorrr{0.783} & \cellcolorrr{0.877} & \cellcolorrr{0.821} & \cellcolorrr{0.719} \\ 
            ~ & ~ & ROUGE-L ($\uparrow$) & \cellcolorrr{0.358} & \cellcolorrr{0.579} & \cellcolorrr{0.709} & \cellcolorrr{0.706} & \cellcolorrr{0.663} \\ \cline{2-8}
            ~ & \multirow{2}{*}{Mid} & BERTScore ($\uparrow$) & \cellcolorrr{0.419} & \cellcolorrr{0.619} & \cellcolorrr{0.791} & \cellcolorrr{0.820} & \cellcolorrr{0.724} \\ 
            ~ & ~ & ROUGE-L ($\uparrow$) & \cellcolorrr{0.310} & \cellcolorrr{0.482} & \cellcolorrr{0.617} & \cellcolorrr{0.661} & \cellcolorrr{0.592} \\ \cline{2-8}
            ~ & \multirow{2}{*}{Last} & BERTScore ($\uparrow$) & \cellcolorrr{0.579} & \cellcolorrr{0.807} & \cellcolorrr{0.894} & \cellcolorrr{0.878} & \cellcolorrr{0.727} \\ 
            ~ & ~ & ROUGE-L ($\uparrow$) & \cellcolorrr{0.455} & \cellcolorrr{0.615} & \cellcolorrr{0.680} & \cellcolorrr{0.648} & \cellcolorrr{0.485} \\ \cline{2-8}
            ~ & \multicolumn{2}{c|}{Average Time(s)} & 1.17 & 2.13 & 5.06 & 12.65 & 21.58 \\ \hline
        \hline
        \end{tabular}
    }
    \vspace{-0.5em}
\end{table}

\partitle{Outlier Detection Threshold}
We experimented with different outlier detection thresholds on the LLaMA-3.2-1B model. The results are shown in Table~\ref{table:gap}. We observed that setting the threshold to $3\sigma_{\text{other}}$ (i.e., treating a value as an outlier if it is more than three standard deviations below the mean of the $d_{\text{other}}$ distribution, which corresponds to approximately 0.13\% of a standard normal distribution) yields the highest reconstruction accuracy. A threshold that is too low (e.g., $2\sigma_{\text{other}}$) leads to misidentifying incorrect tokens as the target, thus reducing accuracy. Conversely, a threshold that is too high (e.g., $4\sigma_{\text{other}}$) can cause the correct token to be missed, which also decreases accuracy while significantly increasing the attack time. Therefore, we set the outlier detection threshold to $3\sigma_{\text{other}}$ for all subsequent experiments.

\begin{table}[t]
    \centering
    \tabcolsep=1.0mm
    \renewcommand{\arraystretch}{1.1}
    \caption{Impact of different batch sizes on reconstruction accuracy, using a fixed outlier detection threshold of $3\sigma_{\text{other}}$.}
    \label{table:batch_size}
    \scalebox{0.79}{
        \begin{tabular}{c|c|c|ccccc}
        \hline
        \hline
            \multirow{2}{*}{Model} & \multirow{2}{*}{Layer} & \multirow{2}{*}{Metric} & \multicolumn{5}{c}{Batch Size} \\ \cline{4-8}
            ~ & ~ & ~ & 64 & 128 & 256 & 512 & 1024 \\ \hline
            \multirow{7}{*}{LLaMA-3.2-1B} & \multirow{2}{*}{First} & BERTScore ($\uparrow$) & \cellcolorrr{0.830} & \cellcolorrr{0.856} & \cellcolorrr{0.877} & \cellcolorrr{0.869} & \cellcolorrr{0.837} \\ 
            ~ & ~ & ROUGE-L ($\uparrow$) & \cellcolorrr{0.711} & \cellcolorrr{0.723} & \cellcolorrr{0.709} & \cellcolorrr{0.666} & \cellcolorrr{0.594} \\ \cline{2-8}
            ~ & \multirow{2}{*}{Mid} & BERTScore ($\uparrow$) & \cellcolorrr{0.765} & \cellcolorrr{0.786} & \cellcolorrr{0.791} & \cellcolorrr{0.771} & \cellcolorrr{0.753} \\ 
            ~ & ~ & ROUGE-L ($\uparrow$) & \cellcolorrr{0.590} & \cellcolorrr{0.617} & \cellcolorrr{0.617} & \cellcolorrr{0.588} & \cellcolorrr{0.554} \\ \cline{2-8}
            ~ & \multirow{2}{*}{Last} & BERTScore ($\uparrow$) & \cellcolorrr{0.837} & \cellcolorrr{0.865} & \cellcolorrr{0.894} & \cellcolorrr{0.902} & \cellcolorrr{0.887} \\ 
            ~ & ~ & ROUGE-L ($\uparrow$) & \cellcolorrr{0.513} & \cellcolorrr{0.616} & \cellcolorrr{0.680} & \cellcolorrr{0.674} & \cellcolorrr{0.636} \\ \cline{2-8}
            ~ & \multicolumn{2}{c|}{Average Time(s)} & 12.94 & 7.58 & 5.06 & 4.04 & 4.62 \\ \hline
        \hline
        \end{tabular}
    }
    \vspace{-0.5em}
\end{table}

\partitle{Batch Size}
We evaluated the effect of different batch sizes on the LLaMA-3.2-1B model. As shown in Table~\ref{table:batch_size}, with the outlier threshold fixed at $3\sigma_{\text{other}}$, a batch size of 256 achieves the highest reconstruction accuracy. Theoretically, a larger batch size may provide a more robust statistical sample of the $d_{\text{other}}$ distances, leading to higher accuracy. However, our experiments show that as the batch size increases beyond 256, the reconstruction accuracy paradoxically decreases. We attribute this to a mismatch between the batch size and the fixed threshold; a larger batch would likely require a higher, more stringent threshold to maintain accuracy. However, larger batches increase GPU memory consumption, and a higher threshold would multiplicatively increase attack time. To balance accuracy, memory usage, and attack speed, we chose a batch size of 256 for our experiments.

\begin{table}[t]
    \centering
    \tabcolsep=1.0mm
    \renewcommand{\arraystretch}{1.1}
    \caption{Impact of different adversarial instructions on Injection Attack success rates.}
    \label{table:injection}
    \scalebox{0.85}{
        \begin{threeparttable}
            \begin{tabular}{c|c|cccc}
            \hline
            \hline
                \multirow{2}{*}{Model} & \multirow{2}{*}{Metric} & \multicolumn{4}{c}{Inject Instruction} \\ \cline{3-6}
                ~ & ~ & \makecell[c]{Ins1}   
        & \makecell[c]{Ins2} & \makecell[c]{Ins3} & \makecell[c]{Ins4} \\ \hline
                \multirow{2}{*}{LLaMA-7B} & BERTScore ($\uparrow$) & \cellcolorrr{0.765} & \cellcolorrr{0.716} & \cellcolorrr{0.557} & \cellcolorrr{0.598} \\ 
                ~ & ROUGE-L ($\uparrow$) & \cellcolorrr{0.687} & \cellcolorrr{0.606} & \cellcolorrr{0.449} & \cellcolorrr{0.473} \\ \hline
                \multirow{2}{*}{LLaMA-3.2-1B} & BERTScore ($\uparrow$) & \cellcolorrr{0.544} & \cellcolorrr{0.533} & \cellcolorrr{0.423} & \cellcolorrr{0.353} \\ 
                ~ & ROUGE-L ($\uparrow$) & \cellcolorrr{0.315} & \cellcolorrr{0.358} & \cellcolorrr{0.232} & \cellcolorrr{0.217} \\ \hline
                \multirow{2}{*}{LLaMA-3.2-3B-Instruct} & BERTScore ($\uparrow$) & \cellcolorrr{0.540} & \cellcolorrr{0.360} & \cellcolorrr{0.506} & \cellcolorrr{0.271} \\ 
                ~ & ROUGE-L ($\uparrow$) & \cellcolorrr{0.324} & \cellcolorrr{0.157} & \cellcolorrr{0.358} & \cellcolorrr{0.124} \\ \hline
                \multirow{2}{*}{LLaMA-3.1-8B} & BERTScore ($\uparrow$) & \cellcolorrr{0.616} & \cellcolorrr{0.544} & \cellcolorrr{0.432} & \cellcolorrr{0.457} \\ 
                ~ & ROUGE-L ($\uparrow$) & \cellcolorrr{0.447} & \cellcolorrr{0.365} & \cellcolorrr{0.275} & \cellcolorrr{0.279} \\ \hline
                \multirow{2}{*}{LLaMA-3.1-8B-Distilled} & BERTScore ($\uparrow$) & \cellcolorrr{0.610} & \cellcolorrr{0.536} & \cellcolorrr{0.348} & \cellcolorrr{0.434} \\ 
                ~ & ROUGE-L ($\uparrow$) & \cellcolorrr{0.421} & \cellcolorrr{0.348} & \cellcolorrr{0.218} & \cellcolorrr{0.249} \\ \hline
                \multirow{2}{*}{Qwen2.5-Math-7B} & BERTScore ($\uparrow$) & \cellcolorrr{0.422} & \cellcolorrr{0.381} & \cellcolorrr{0.413} & \cellcolorrr{0.329} \\ 
                ~ & ROUGE-L ($\uparrow$) & \cellcolorrr{0.286} & \cellcolorrr{0.222} & \cellcolorrr{0.281} & \cellcolorrr{0.194} \\ \hline
            \hline
            \end{tabular}

            \begin{tablenotes}
                \item \textit{Note:} The specific instructions are: ``Repeat the previous content.'' (Ins1), ``Summarize the previous content.'' (Ins2), ``Repeat what I said.'' (Ins3), and ``Summarize what I said.'' (Ins4).
            \end{tablenotes}
        \end{threeparttable}
    }
    \vspace{-0.5em}
\end{table}

\subsubsection{Ablation Study of Injection Attack}
\label{app:injection_ablation}
We tested various instructions against each model's KV-cache, with the results shown in Table~\ref{table:injection}. The instruction ``Repeat the previous content.'' achieved the highest overall reconstruction accuracy across all models, with an average BERTScore of 0.58 and ROUGE-L of 0.42. 

\subsubsection{Attack Generalization Across Datasets}
\label{app:different_datasets}

\begin{table*}[t!]
    \centering
    \tabcolsep=2.5mm
    \renewcommand{\arraystretch}{1.1}
    \caption{Attack generalization across datasets on LLaMA-3.2-1B.}
    \label{table:different_datasets}
    \scalebox{0.80}{
        \begin{tabular}{c|c|c|c|ccc|ccc|c}
        \hline
        \hline
            \multirow{2}{*}{Model} & \multirow{2}{*}{Datasets} & \multirow{2}{*}{Metric} & Inversion & \multicolumn{3}{c|}{Collision} & \multicolumn{3}{c|}{Collision+} & Injection \\ \cline{4-11}
            ~ & ~ & ~ & First & First & Mid & Last & First & Mid & Last & All \\ \hline
            \multirow{6}{*}{LLaMA-3.2-1B} & \multirow{2}{*}{Alpaca} & BERTScore ($\uparrow$) & \cellcolorrr{1.000} & \cellcolorrr{0.942} & \cellcolorrr{0.910} & \cellcolorrr{0.936} & \cellcolorrr{1.000} & \cellcolorrr{1.000} & \cellcolorrr{1.000} & \cellcolorrr{0.570} \\ 
            ~ & ~ & ROUGE-L ($\uparrow$) & \cellcolorrr{1.000} & \cellcolorrr{0.897} & \cellcolorrr{0.700} & \cellcolorrr{0.923} & \cellcolorrr{1.000} & \cellcolorrr{1.000} & \cellcolorrr{1.000} & \cellcolorrr{0.374} \\ \cline{2-11}
            ~ & \multirow{2}{*}{GSM8K} & BERTScore ($\uparrow$) & \cellcolorrr{1.000} & \cellcolorrr{0.890} & \cellcolorrr{0.716} & \cellcolorrr{0.911} & \cellcolorrr{1.000} & \cellcolorrr{1.000} & \cellcolorrr{1.000} & \cellcolorrr{0.632} \\ 
            ~ & ~ & ROUGE-L ($\uparrow$) & \cellcolorrr{1.000} & \cellcolorrr{0.784} & \cellcolorrr{0.493} & \cellcolorrr{0.800} & \cellcolorrr{1.000} & \cellcolorrr{0.999} & \cellcolorrr{1.000} & \cellcolorrr{0.496} \\ \cline{2-11}
            ~ & \multirow{2}{*}{LMSYS-Chat-1M} & BERTScore ($\uparrow$) & \cellcolorrr{1.000} & \cellcolorrr{0.877} & \cellcolorrr{0.791} & \cellcolorrr{0.894} & \cellcolorrr{1.000} & \cellcolorrr{1.000} & \cellcolorrr{1.000} & \cellcolorrr{0.544} \\ 
            ~ & ~ & ROUGE-L ($\uparrow$) & \cellcolorrr{0.994} & \cellcolorrr{0.709} & \cellcolorrr{0.617} & \cellcolorrr{0.680} & \cellcolorrr{0.994} & \cellcolorrr{0.994} & \cellcolorrr{0.994} & \cellcolorrr{0.315} \\ \hline
        \hline
        \end{tabular}
    }
    \vspace{-0.5em}
\end{table*}

To validate the generalization capability of proposed attack, we also conducted evaluations on two datasets from different domains: Alpaca~\cite{taori2023stanford} (instruction-following) and GSM8K~\cite{cobbe2021gsm8k} (mathematical reasoning).
As shown in Table~\ref{table:different_datasets}, our attacks achieved high reconstruction accuracy across all tested datasets.
This result confirms that the effectiveness is not confined to specific data distributions or task types, but is instead highly generalizable.

\subsection{DP Baseline Parameter Selection} 
\label{app:dp_ablation}
To establish a robust DP baseline, we first defined its parameterization methodology. We then conducted experiments to select a configuration that balances utility and privacy for comparison against KV-Cloak.

\begin{itemize}[leftmargin=*]
    \item \textbf{Noise Application:} As illustrated in Figure~\ref{fig:kvcache_distribution}, the element distributions of the K and V caches differ significantly. Consequently, we apply $(\epsilon, \delta)$-DP Gaussian noise to the K and V tensors independently.
    
    \item \textbf{Clipping Threshold $C$:} The noise magnitude in DP is determined by the function's sensitivity, which we control by clipping the Frobenius norm of the K and V tensors. To find an appropriate clipping threshold, we generated 1,000 long sequences (approximately 2,000 tokens each) from the MMLU dataset, recorded the distribution of the resulting KV-cache Frobenius norms, and experimented with thresholds corresponding to different percentiles of this distribution.
    
    \item \textbf{Privacy Budget $\epsilon$:} This parameter governs the fundamental trade-off between privacy and utility. We evaluated a wide range of $\epsilon$ values to map out their impact on model accuracy.
    
    \item \textbf{Failure Probability $\delta$:} This represents the probability of the privacy guarantee being violated. We adopt the common standard value of $\delta = 10^{-5}$ for all DP experiments.
\end{itemize}

\begin{table}[t]
    \centering
    \tabcolsep=1.0mm
    \renewcommand{\arraystretch}{1.1}
    \caption{Privacy-utility trade-off: LLaMA-3.2-1B inference accuracy under DP with varying noise ($\epsilon$) and clipping thresholds. The 50th percentile norm is highlighted as the baseline for subsequent comparisons.}
    \label{table:dp_accuracy}
    \scalebox{0.77}{
        \begin{tabular}{c|c|c|ccccc}
        \hline
        \hline
            \multirow{2}{*}{Model} & \multirow{2}{*}{Norm Ratio} & \multirow{2}{*}{Metric} & \multicolumn{5}{c}{$\epsilon$} \\ \cline{4-8}
            ~ & ~ & ~ & $1$ & $10$ & $10^{7}$ & $10^{8}$ & $10^{9}$ \\ \hline
            \multirow{6}{*}{LLaMA-3.2-1B} & \multirow{2}{*}{50\%} & MMLU ($\uparrow$) & 0.051 & 0.052 & 0.052 & 0.262 & 0.299 \\ 
            ~ & ~ & SQuAD ($\uparrow$) & 0.000 & 0.000 & 0.000 & 0.258 & 0.443 \\ \cline{2-8}
            ~ & \multirow{2}{*}{90\%} & MMLU ($\uparrow$) & 0.053 & 0.053 & 0.045 & 0.259 & 0.309 \\ 
            ~ & ~ & SQuAD ($\uparrow$) & 0.000 & 0.000 & 0.000 & 0.171 & 0.435 \\ \cline{2-8}
            ~ & \multirow{2}{*}{95\%} & MMLU ($\uparrow$) & 0.052 & 0.053 & 0.045 & 0.252 & 0.309 \\ 
            ~ & ~ & SQuAD ($\uparrow$) & 0.000 & 0.000 & 0.000 & 0.136 & 0.437 \\ \hline
        \hline
        \end{tabular}
    }
    \vspace{-0.5em}
\end{table}

\begin{table}[t]
    \centering
    \tabcolsep=0.8mm
    \renewcommand{\arraystretch}{1.1}
    \caption{Effectiveness of the Inversion, Collision, and Injection attacks against different layers of the KV-cache from the LLaMA-3.2-1B model, under various DP mechanisms.}
    \label{table:dp_security}
    \scalebox{0.67}{
        \begin{tabular}{c|c|c|c|ccc|c}
        \hline
        \hline
            \multirow{2}{*}{Model} & \multirow{2}{*}{Protect Type} & \multirow{2}{*}{Metric} & Inversion & \multicolumn{3}{c|}{Collision} & Injection \\ \cline{4-8}
            ~ & ~ & ~ & First & First & Mid & Last & All \\ \hline
            \multirow{8}{*}{LLaMA-3.2-1B} & \multirow{2}{*}{Plaintext} & BERTScore ($\downarrow$) & \cellcolorrr{1.000} & \cellcolorrr{0.877} & \cellcolorrr{0.791} & \cellcolorrr{0.894} & \cellcolorrr{0.544} \\ 
            ~ & ~ & ROUGE-L ($\downarrow$) & \cellcolorrr{0.994} & \cellcolorrr{0.709} & \cellcolorrr{0.617} & \cellcolorrr{0.680} & \cellcolorrr{0.315} \\ \cline{2-8}
            ~ & \multirow{2}{*}{($10^{7},10^{-5}$)-DP} & BERTScore ($\downarrow$) & \cellcolorrr{0.096} & \cellcolorrr{0.469} & \cellcolorrr{0.651} & \cellcolorrr{0.672} & \cellcolorrr{0.131} \\ 
            ~ & ~ & ROUGE-L ($\downarrow$) & \cellcolorrr{0.073} & \cellcolorrr{0.353} & \cellcolorrr{0.402} & \cellcolorrr{0.336} & \cellcolorrr{0.067} \\ \cline{2-8}
            ~ & \multirow{2}{*}{($10^{8},10^{-5}$)-DP} & BERTScore ($\downarrow$) & \cellcolorrr{0.633} & \cellcolorrr{0.849} & \cellcolorrr{0.763} & \cellcolorrr{0.849} & \cellcolorrr{0.393} \\ 
            ~ & ~ & ROUGE-L ($\downarrow$) & \cellcolorrr{0.622} & \cellcolorrr{0.604} & \cellcolorrr{0.587} & \cellcolorrr{0.604} & \cellcolorrr{0.248} \\ \cline{2-8}
            ~ & \multirow{2}{*}{($10^{9},10^{-5}$)-DP} & BERTScore ($\downarrow$) & \cellcolorrr{0.994} & \cellcolorrr{0.808} & \cellcolorrr{0.786} & \cellcolorrr{0.886} & \cellcolorrr{0.524} \\ 
            ~ & ~ & ROUGE-L ($\downarrow$) & \cellcolorrr{0.980} & \cellcolorrr{0.635} & \cellcolorrr{0.610} & \cellcolorrr{0.667} & \cellcolorrr{0.304} \\ \hline
        \hline
        \end{tabular}
    }
    \vspace{-0.5em}
\end{table}

Our experimental results, presented in Table~\ref{table:dp_accuracy}, reveal a stark trade-off between privacy and model accuracy for the DP baseline. Under conventionally strong privacy settings (e.g., $\epsilon=1$ or $\epsilon=10$), model accuracy on both MMLU and SQuAD collapses to the level of random guessing, regardless of the chosen clipping threshold. Accuracy only begins to recover when the privacy budget is substantially relaxed: at $\epsilon=10^8$, it reaches 59.13\% of the unprotected baseline's accuracy; and at $\epsilon=10^9$, it improves to 93.61\%. This extreme sensitivity is due to the highly sparse nature of the KV-cache, where most elements are near zero. Directly adding noise disproportionately perturbs the cache's delicate structure, severely degrading model performance unless the noise is made negligible by an extremely large $\epsilon$. And its defensive efficacy, presented in Table~\ref{table:dp_security}, is strongly correlated with the privacy budget $\epsilon$. With a weak budget of $\epsilon=10^8$, the accuracy of the Inversion and Injection attacks is reduced, but the Collision Attack can still recover some useful information. As $\epsilon$ is strengthened to $10^7$, the overall attack success rate decreases further. However, the Collision Attack can still achieve a reconstruction with over 55\% semantic similarity. Importantly, this protection comes at the cost of model accuracy, a trade-off we will discuss in detail in the next section.

To balance security and accuracy for our comparative analysis, we selected $\epsilon=10^8$ and a clipping norm at the 50th percentile for subsequent experiments, as this offered a reasonable degree of utility for the DP baseline.

\subsection{Evaluation of Security on the Remaining Models}
\label{app:defense_evaluation_other}

\begin{table*}[t]
    \centering
    \tabcolsep=2.5mm
    \renewcommand{\arraystretch}{1.1}
    \caption{Efficacy of Defense Mechanisms Against Input Reconstruction Attacks on the Remaining Models.}
    \label{table:defense_evaluation_other}
    \scalebox{0.84}{
        \begin{tabular}{c|c|c|c|ccc|ccc|c}
        \hline
        \hline
            \multirow{2}{*}{Model} & \multirow{2}{*}{Protect Type} & \multirow{2}{*}{Metric} & Inversion & \multicolumn{3}{c|}{Collision} & \multicolumn{3}{c|}{Collision+} & Injection \\ \cline{4-11}
            ~ & ~ & ~ & First & First & Mid & Last & First & Mid & Last & All \\ \hline
            \multirow{6}{*}{LLaMA-3.2-3B-Instruct} & \multirow{2}{*}{Original} & BERTScore ($\downarrow$)& \cellcolorrr{0.055} & \cellcolorrr{0.782} & \cellcolorrr{0.668} & \cellcolorrr{0.820} & \cellcolorrr{1.000} & \cellcolorrr{1.000} & \cellcolorrr{1.000} & \cellcolorrr{0.540} \\ 
            ~ & ~ & ROUGE-L ($\downarrow$)& \cellcolorrr{0.000} & \cellcolorrr{0.732} & \cellcolorrr{0.456} & \cellcolorrr{0.621} & \cellcolorrr{0.994} & \cellcolorrr{0.994} & \cellcolorrr{0.994} & \cellcolorrr{0.324} \\ \cline{2-11}
            ~ & \multirow{2}{*}{KV-Cloak} & BERTScore ($\downarrow$)& \cellcolorrr{0.088} & \cellcolorrr{0.069} & \cellcolorrr{0.070} & \cellcolorrr{0.069} & \cellcolorrr{0.033} & \cellcolorrr{0.033} & \cellcolorrr{0.033} & \cellcolorrr{0.088} \\ 
            ~ & ~ & ROUGE-L ($\downarrow$)& \cellcolorrr{0.000} & \cellcolorrr{0.000} & \cellcolorrr{0.000} & \cellcolorrr{0.000} & \cellcolorrr{0.042} & \cellcolorrr{0.042} & \cellcolorrr{0.042} & \cellcolorrr{0.000} \\ \cline{2-11}
            ~ & \multirow{2}{*}{($10^{8},10^{-5}$)-DP} & BERTScore ($\downarrow$)& \cellcolorrr{0.061} & \cellcolorrr{0.223} & \cellcolorrr{0.592} & \cellcolorrr{0.760} & \cellcolorrr{0.967} & \cellcolorrr{0.938} & \cellcolorrr{1.000} & \cellcolorrr{0.129} \\ 
            ~ & ~ & ROUGE-L ($\downarrow$)& \cellcolorrr{0.000} & \cellcolorrr{0.261} & \cellcolorrr{0.360} & \cellcolorrr{0.517} & \cellcolorrr{0.951} & \cellcolorrr{0.907} & \cellcolorrr{0.994} & \cellcolorrr{0.032} \\ \hline
            \multirow{6}{*}{LLaMA-3.1-8B} & \multirow{2}{*}{Original} & BERTScore ($\downarrow$)& \cellcolorrr{0.071} & \cellcolorrr{0.873} & \cellcolorrr{0.652} & \cellcolorrr{0.764} & \cellcolorrr{1.000} & \cellcolorrr{1.000} & \cellcolorrr{1.000} & \cellcolorrr{0.616} \\ 
            ~ & ~ & ROUGE-L ($\downarrow$)& \cellcolorrr{0.000} & \cellcolorrr{0.825} & \cellcolorrr{0.443} & \cellcolorrr{0.564} & \cellcolorrr{0.994} & \cellcolorrr{0.994} & \cellcolorrr{0.994} & \cellcolorrr{0.447} \\ \cline{2-11}
            ~ & \multirow{2}{*}{KV-Cloak} & BERTScore ($\downarrow$)& \cellcolorrr{0.076} & \cellcolorrr{0.069} & \cellcolorrr{0.069} & \cellcolorrr{0.069} & \cellcolorrr{0.041} & \cellcolorrr{0.041} & \cellcolorrr{0.041} & \cellcolorrr{0.084} \\ 
            ~ & ~ & ROUGE-L ($\downarrow$)& \cellcolorrr{0.004} & \cellcolorrr{0.000} & \cellcolorrr{0.000} & \cellcolorrr{0.000} & \cellcolorrr{0.003} & \cellcolorrr{0.003} & \cellcolorrr{0.003} & \cellcolorrr{0.000} \\ \cline{2-11}
            ~ & \multirow{2}{*}{($10^{8},10^{-5}$)-DP} & BERTScore ($\downarrow$)& \cellcolorrr{0.076} & \cellcolorrr{0.343} & \cellcolorrr{0.526} & \cellcolorrr{0.614} & \cellcolorrr{0.639} & \cellcolorrr{0.986} & \cellcolorrr{0.999} & \cellcolorrr{0.115} \\ 
            ~ & ~ & ROUGE-L ($\downarrow$)& \cellcolorrr{0.003} & \cellcolorrr{0.328} & \cellcolorrr{0.284} & \cellcolorrr{0.419} & \cellcolorrr{0.639} & \cellcolorrr{0.947} & \cellcolorrr{0.994} & \cellcolorrr{0.057} \\ \hline
            \multirow{6}{*}{Qwen2.5-Math-7B} & \multirow{2}{*}{Original} & BERTScore ($\downarrow$)& \cellcolorrr{0.229} & \cellcolorrr{0.918} & \cellcolorrr{0.552} & \cellcolorrr{0.783} & \cellcolorrr{1.000} & \cellcolorrr{0.983} & \cellcolorrr{0.996} & \cellcolorrr{0.422} \\ 
            ~ & ~ & ROUGE-L ($\downarrow$)& \cellcolorrr{0.186} & \cellcolorrr{0.842} & \cellcolorrr{0.355} & \cellcolorrr{0.580} & \cellcolorrr{1.000} & \cellcolorrr{0.977} & \cellcolorrr{0.996} & \cellcolorrr{0.286} \\ \cline{2-11}
            ~ & \multirow{2}{*}{KV-Cloak} & BERTScore ($\downarrow$)& \cellcolorrr{0.099} & \cellcolorrr{0.069} & \cellcolorrr{0.069} & \cellcolorrr{0.070} & \cellcolorrr{0.112} & \cellcolorrr{0.112} & \cellcolorrr{0.113} & \cellcolorrr{0.075} \\ 
            ~ & ~ & ROUGE-L ($\downarrow$)& \cellcolorrr{0.011} & \cellcolorrr{0.000} & \cellcolorrr{0.000} & \cellcolorrr{0.000} & \cellcolorrr{0.000} & \cellcolorrr{0.000} & \cellcolorrr{0.000} & \cellcolorrr{0.000} \\ \cline{2-11}
            ~ & \multirow{2}{*}{($10^{8},10^{-5}$)-DP} & BERTScore ($\downarrow$)& \cellcolorrr{0.108} & \cellcolorrr{0.879} & \cellcolorrr{0.274} & \cellcolorrr{0.317} & \cellcolorrr{0.331} & \cellcolorrr{0.432} & \cellcolorrr{0.373} & \cellcolorrr{0.325} \\ 
            ~ & ~ & ROUGE-L ($\downarrow$)& \cellcolorrr{0.018} & \cellcolorrr{0.790} & \cellcolorrr{0.100} & \cellcolorrr{0.143} & \cellcolorrr{0.336} & \cellcolorrr{0.445} & \cellcolorrr{0.404} & \cellcolorrr{0.208} \\ \hline
        \hline
        \end{tabular}
    }
    \vspace{-0.5em}
\end{table*}

As shown in Table~\ref{table:defense_evaluation_other}, KV-Cloak completely thwarts all our proposed attacks, reducing the quality of any reconstructed text to a level statistically indistinguishable from random noise.

\subsection{Performance Analysis and the Impact of Operator Fusion}
\label{sec:performance_analysis}

A critical aspect of any practical defense mechanism is its performance overhead. In this section, we analyze the computational cost of KV-Cloak and demonstrate the significant efficiency gains achieved through our operator fusion technique.

\partitle{Overhead of a Naive Implementation}
Without operator fusion, a naive implementation would apply the obfuscation transform $K' = S\hat{P}(K+A)M$ and its inverse as explicit. Neglecting the computationally inexpensive matrix additions involving $A$, the primary overhead stems from matrix multiplications. For a single KV-cache block of size $b \times d$:
\begin{itemize}[leftmargin=*]
    \item The obfuscation operation requires approximately $b^3$ (for $S\hat{P}$), $b^2d$ (for $(S\hat{P})K$), and $bd^2$ (for $(S\hat{P}K)M$) floating-point multiplications.
    \item The de-obfuscation requires an additional $b^2d$ (for $S^{-1}K'$) and $bd^2$ (for $(K')M^{-1}$) multiplications.
\end{itemize}
This results in a total of approximately $b^3 + 2b^2d + 2bd^2$ multiplications per block per decoding step. To put this into perspective, the cost of re-computing the same KV-cache block from the LLM's hidden states (dimension $D$) is $b \cdot D \cdot d$. For a model like LLaMA-3.1-8B (with $b=16, d=128, D=4096$), the naive obfuscation overhead constitutes a substantial 7.1\% of the re-computation cost.

\partitle{Efficiency Gains from Operator Fusion}
By fusing the matrix $M$ and its inverse into the model's weights offline, as described in Section~\ref{sec:operator_fusion}, we eliminate the two most expensive online multiplications ($bd^2$ terms). The online obfuscation and de-obfuscation, governed by Eq.~(\ref{eq:obf_SPA}), now only require approximately $b^3 + 2b^2d$ multiplications.

Revisiting the LLaMA-3.1-8B example, this optimization reduces the computational overhead to just 0.83\% of the re-computation cost. This represents a nearly 8-fold reduction in latency compared to the naive implementation (specifically, the new cost is 11.72\% of the original overhead). This dramatic improvement makes the runtime performance impact of KV-Cloak minimal and highly practical for real-world deployment.

\partitle{Auxiliary Costs}
Our analysis primarily focuses on floating-point multiplications, which dominate the computational cost. However, we acknowledge other minor costs, such as the generation of the one-time permutation matrix $\hat{P}$, the element-wise additions for the mask $A$, and function call overhead. These costs are considered secondary for several reasons: the generation of $\hat{P}$ can be performed asynchronously in parallel with other computations; matrix addition has a much lower complexity than multiplication; and any remaining overhead can be further optimized through techniques like computation graph optimization and hardware acceleration.

\subsection{Architectural Compatibility with PagedAttention}
The compatibility of KV-Cloak with modern inference engines stems from its core ``block-oriented'' design principle. All cryptographic operations—both obfuscation and de-obfuscation—are self-contained within a single physical memory block. This design intentionally creates no cross-block dependencies, allowing a memory manager like vLLM to schedule, copy, swap, and share physical blocks freely, without any awareness of their obfuscated contents.

\begin{table}[t]
    \centering
    \tabcolsep=1.0mm
    \renewcommand{\arraystretch}{1.1}
    \caption{Impact of PagedAttention block size on LLaMA-3.2-1B inference accuracy under KV-Cloak protection.}
    \label{table:block_size_accuracy}
    \scalebox{0.9}{
        \begin{tabular}{c|c|c|ccc}
        \hline
        \hline
            \multirow{2}{*}{Model} & \multirow{2}{*}{Metric} & \multirow{2}{*}{Plaintext} & \multicolumn{3}{c}{Block Size} \\ \cline{4-6}
            ~ & ~ & ~ & 16 & 32 & 64 \\ \hline
            \multirow{2}{*}{LLaMA-3.2-1B} & MMLU ($\uparrow$) & 0.335 & 0.335 & 0.335 & 0.335 \\ 
            ~ & SQuAD ($\uparrow$) & 0.457 & 0.463 & 0.462 & 0.460 \\ \hline
        \hline
        \end{tabular}
        \vspace{-1.0em}
    }
\end{table}

\begin{table}[t]
    \centering
    \tabcolsep=1.0mm
    \renewcommand{\arraystretch}{1.1}
    \caption{Computational overhead of KV-Cloak (Fused vs. No Fuse) across different PagedAttention block sizes on LLaMA-3.1-8B.}
    \label{table:block_size_time}
    \scalebox{0.9}{
        \begin{tabular}{c|c|c|ccc}
        \hline
        \hline
            \multirow{2}{*}{Model} & \multirow{2}{*}{Prefill} & \multirow{2}{*}{Type}& \multicolumn{3}{c}{Block Size} \\ \cline{4-6}
            ~ & ~ & ~ & 16 & 32 & 64 \\ \hline
            \multirow{4}{*}{LLaMA-3.1-8B} & \multirow{4}{*}{3438.5} 
            & \multirow{2}{*}{No Fuse} & 52.73 & 45.45 & 28.60 \\ 
            ~ & ~ & ~ & +1.53\% & +1.32\% & +0.83\% \\ \cline{3-6}
            ~ & ~ & \multirow{2}{*}{Fused} & 15.41 & 10.17 & 12.33 \\ 
            ~ & ~ & ~ & +0.45\% & +0.30\% & +0.36\% \\ \hline
        \hline
        \end{tabular}
    }
    \vspace{-1.0em}
\end{table}

To empirically validate this compatibility, we evaluated KV-Cloak with the most common PagedAttention block sizes: 16, 32, and 64. We measured accuracy on LLaMA-3.2-1B (Table~\ref{table:block_size_accuracy}) and latency on LLaMA-3.1-8B (Table~\ref{table:block_size_time}). As shown in Table~\ref{table:block_size_accuracy}, KV-Cloak is virtually lossless. It preserves the baseline MMLU accuracy perfectly and results in negligible, statistically insignificant variations on SQuAD. In terms of performance (Table~\ref{table:block_size_time}), the overhead of the optimized, fused implementation remains consistently low, adding only $< 0.45\%$ latency relative to the prefill computation across all block sizes.

These results confirm that KV-Cloak's design has no fundamental conflicts with the PagedAttention memory management model. Its negligible impact on accuracy and its low, stable overhead across various block sizes demonstrate that a full integration into an inference engine like vLLM is a practical and feasible engineering task.

\subsection{Broader Impact of KV-Cloak for LLM Inference Security}

Although KV-Cloak targets the KV-cache specifically, it underscores a broader issue: the internal states of large language models represent a rich and vulnerable attack surface. As models grow in scale and architectural complexity (e.g., via Mixture-of-Experts~\cite{shazeer2017sgmoe}), they produce substantial context-dependent intermediate data—such as activations and attention weights—that may leak sensitive information. KV-Cloak introduces a lightweight, structure-aware obfuscation approach as an alternative to costly cryptographic methods. By exploiting mathematical reversibility, it preserves model accuracy while embedding sufficient algebraic complexity to resist cryptanalysis. This algorithm-architecture co-design paradigm offers a promising direction for enhancing LLM inference security.


\begin{thebibliography}{10}
\providecommand{\url}[1]{#1}
\csname url@samestyle\endcsname
\providecommand{\newblock}{\relax}
\providecommand{\bibinfo}[2]{#2}
\providecommand{\BIBentrySTDinterwordspacing}{\spaceskip=0pt\relax}
\providecommand{\BIBentryALTinterwordstretchfactor}{4}
\providecommand{\BIBentryALTinterwordspacing}{\spaceskip=\fontdimen2\font plus
\BIBentryALTinterwordstretchfactor\fontdimen3\font minus \fontdimen4\font\relax}
\providecommand{\BIBforeignlanguage}[2]{{%
\expandafter\ifx\csname l@#1\endcsname\relax
\typeout{** WARNING: IEEEtranS.bst: No hyphenation pattern has been}%
\typeout{** loaded for the language `#1'. Using the pattern for}%
\typeout{** the default language instead.}%
\else
\language=\csname l@#1\endcsname
\fi
#2}}
\providecommand{\BIBdecl}{\relax}
\BIBdecl

\bibitem{abadi2016deep}
M.~Abadi, A.~Chu, I.~Goodfellow, H.~B. McMahan, I.~Mironov, K.~Talwar, and L.~Zhang, ``Deep learning with differential privacy,'' in \emph{ACM SIGSAC Conference on Computer and Communications Security}, 2016.

\bibitem{acar2018survey}
A.~Acar, H.~Aksu, A.~S. Uluagac, and M.~Conti, ``A survey on homomorphic encryption schemes: Theory and implementation,'' \emph{ACM Computing Surveys}, vol.~51, no.~4, pp. 1--35, 2018.

\bibitem{ainslie2023gqa}
J.~Ainslie, J.~Lee-Thorp, M.~de~Jong, Y.~Zemlyanskiy, F.~Lebr{\'o}n, and S.~Sanghai, ``Gqa: Training generalized multi-query transformer models from multi-head checkpoints,'' \emph{arXiv preprint arXiv:2305.13245}, 2023.

\bibitem{alenezi2020symmetric}
M.~N. Alenezi, H.~Alabdulrazzaq, and N.~Q. Mohammad, ``Symmetric encryption algorithms: Review and evaluation study,'' \emph{International Journal of Communication Networks and Information Security}, vol.~12, no.~2, pp. 256--272, 2020.

\bibitem{apple2024pcc}
\BIBentryALTinterwordspacing
APPLE. (2024) Private cloud compute: A new frontier for ai privacy in the cloud. [Online]. Available: \url{https://security.apple.com/blog/private-cloud-compute/}
\BIBentrySTDinterwordspacing

\bibitem{bengio2003neural}
Y.~Bengio, R.~Ducharme, P.~Vincent, and C.~Jauvin, ``A neural probabilistic language model,'' \emph{Journal of Machine Learning Research}, vol.~3, no. Feb, pp. 1137--1155, 2003.

\bibitem{carlini2021extracting}
N.~Carlini, F.~Tramer, E.~Wallace, M.~Jagielski, A.~Herbert-Voss, K.~Lee, A.~Roberts, T.~Brown, D.~Song, U.~Erlingsson \emph{et~al.}, ``Extracting training data from large language models,'' in \emph{USENIX Security Symposium}, 2021, pp. 2633--2650.

\bibitem{chen2022teacher}
Y.~Chen, C.~Shen, C.~Wang, and Y.~Zhang, ``Teacher model fingerprinting attacks against transfer learning,'' in \emph{USENIX Security Symposium}, 2022.

\bibitem{cobbe2021gsm8k}
K.~Cobbe, V.~Kosaraju, M.~Bavarian, M.~Chen, H.~Jun, L.~Kaiser, M.~Plappert, J.~Tworek, J.~Hilton, R.~Nakano, C.~Hesse, and J.~Schulman, ``Training verifiers to solve math word problems,'' \emph{arXiv preprint arXiv:2110.14168}, 2021.

\bibitem{dhanuskodi2023creating}
G.~Dhanuskodi, S.~Guha, V.~Krishnan, A.~Manjunatha, R.~Nertney, M.~O'Connor, and P.~Rogers, ``Creating the first confidential gpus,'' \emph{Communications of the ACM}, vol.~67, no.~1, pp. 60--67, 2023.

\bibitem{Dwork2006Calibrating}
C.~Dwork, F.~McSherry, K.~Nissim, and A.~Smith, ``Calibrating noise to sensitivity in private data analysis,'' in \emph{Theory of Cryptography Conference}.\hskip 1em plus 0.5em minus 0.4em\relax Springer Berlin Heidelberg, 2006, pp. 265--284.

\bibitem{grattafiori2024llama}
A.~Grattafiori, A.~Dubey, A.~Jauhri, A.~Pandey, A.~Kadian, A.~Al-Dahle, A.~Letman, A.~Mathur, A.~Schelten, A.~Vaughan \emph{et~al.}, ``The llama 3 herd of models,'' \emph{arXiv preprint arXiv:2407.21783}, 2024.

\bibitem{guo2025deepseek}
D.~Guo, D.~Yang, H.~Zhang, J.~Song, R.~Zhang, R.~Xu, Q.~Zhu, S.~Ma, P.~Wang \emph{et~al.}, ``Deepseek-r1: Incentivizing reasoning capability in llms via reinforcement learning,'' \emph{arXiv preprint arXiv:2501.12948}, 2025.

\bibitem{he2025benchmarking}
Y.~He, H.~She, X.~Qian, X.~Zheng, Z.~Chen, Z.~Qin, and L.~Cavallaro, ``On benchmarking code llms for android malware analysis,'' in \emph{ACM SIGSOFT International Symposium on Software Testing and Analysis Workshop}, 2025.

\bibitem{hendrycks2021ethics}
D.~Hendrycks, C.~Burns, S.~Basart, A.~Critch, J.~Li, D.~Song, and J.~Steinhardt, ``Aligning ai with shared human values,'' \emph{International Conference on Learning Representations}, 2021.

\bibitem{hendryckstest2021}
D.~Hendrycks, C.~Burns, S.~Basart, A.~Zou, M.~Mazeika, D.~Song, and J.~Steinhardt, ``Measuring massive multitask language understanding,'' in \emph{International Conference on Learning Representations}, 2021.

\bibitem{Kwon2023EfficientMM}
W.~Kwon, Z.~Li, S.~Zhuang, Y.~Sheng, L.~Zheng, C.~H. Yu, J.~E. Gonzalez, H.~Zhang, and I.~Stoica, ``Efficient memory management for large language model serving with pagedattention,'' in \emph{Symposium on Operating Systems Principles}, 2023.

\bibitem{li2025every}
A.~Li, B.~Liu, B.~Hu, B.~Li, B.~Zeng, B.~Ye, C.~Tang, C.~Tian, C.~Huang, C.~Zhang \emph{et~al.}, ``Every activation boosted: Scaling general reasoner to 1 trillion open language foundation,'' \emph{arXiv preprint arXiv:2510.22115}, 2025.

\bibitem{li2023sentenceembeddingleaksinformation}
H.~Li, M.~Xu, and Y.~Song, ``Sentence embedding leaks more information than you expect: Generative embedding inversion attack to recover the whole sentence,'' in \emph{Findings of the Association for Computational Linguistics: ACL 2023}, 2023, pp. 14\,022--14\,040.

\bibitem{li2025delay}
X.~Li, Z.~Qin, K.~Ren, C.~Gong, S.~Feng, Y.~Hong, and T.~Wang, ``Delay-allowed differentially private data stream release.'' in \emph{Network and Distributed System Security Symposium}, 2025.

\bibitem{li2025rethinking}
Y.~Li, S.~Shao, Y.~He, J.~Guo, T.~Zhang, Z.~Qin, P.-Y. Chen, M.~Backes, P.~Torr, D.~Tao, and K.~Ren, ``Rethinking data protection in the (generative) artificial intelligence era,'' \emph{arXiv preprint arXiv:2507.03034}, 2025.

\bibitem{lin2024infinite}
B.~Lin, C.~Zhang, T.~Peng, H.~Zhao, W.~Xiao, M.~Sun, A.~Liu, Z.~Zhang, L.~Li, X.~Qiu \emph{et~al.}, ``Infinite-llm: Efficient llm service for long context with distattention and distributed kvcache,'' \emph{arXiv preprint arXiv:2401.02669}, 2024.

\bibitem{lin-2004-rouge}
C.-Y. Lin, ``{ROUGE}: A package for automatic evaluation of summaries,'' in \emph{Text Summarization Branches Out}, 2004, pp. 74--81.

\bibitem{liu2024deepseek}
A.~Liu, B.~Feng, B.~Xue, B.~Wang, B.~Wu, C.~Lu, C.~Zhao, C.~Deng, C.~Zhang, C.~Ruan \emph{et~al.}, ``Deepseek-v3 technical report,'' \emph{arXiv preprint arXiv:2412.19437}, 2024.

\bibitem{morris2023textembeddingsrevealalmost}
J.~Morris, V.~Kuleshov, V.~Shmatikov, and A.~M. Rush, ``Text embeddings reveal (almost) as much as text,'' in \emph{Conference on Empirical Methods in Natural Language Processing}, 2023, pp. 12\,448--12\,460.

\bibitem{paillier1999public}
P.~Paillier, ``Public-key cryptosystems based on composite degree residuosity classes,'' in \emph{International Conference on the Theory and Applications of Cryptographic Techniques}.\hskip 1em plus 0.5em minus 0.4em\relax Springer, 1999, pp. 223--238.

\bibitem{pasquini2025llmmapfingerprintinglargelanguage}
D.~Pasquini, E.~M. Kornaropoulos, and G.~Ateniese, ``Llmmap: Fingerprinting for large language models,'' in \emph{USENIX Security Symposium}, 2025.

\bibitem{pope2023efficiently}
R.~Pope, S.~Douglas, A.~Chowdhery, J.~Devlin, J.~Bradbury, J.~Heek, K.~Xiao, S.~Agrawal, and J.~Dean, ``Efficiently scaling transformer inference,'' \emph{Proceedings of Machine Learning and Systems}, vol.~5, pp. 606--624, 2023.

\bibitem{rajpurkar-etal-2016-squad}
P.~Rajpurkar, J.~Zhang, K.~Lopyrev, and P.~Liang, ``{SQ}u{AD}: 100,000+ questions for machine comprehension of text,'' in \emph{Conference on Empirical Methods in Natural Language Processing}, 2016.

\bibitem{Shao2024DeepSeekV2AS}
Z.~Shao, D.~Dai, D.~Guo, B.~L.~B. Liu), Z.~Wang, and H.~Xin, ``Deepseek-v2: A strong, economical, and efficient mixture-of-experts language model,'' \emph{ArXiv}, 2024.

\bibitem{shazeer2017sgmoe}
N.~Shazeer, A.~Mirhoseini, K.~Maziarz, A.~Davis, Q.~Le, G.~Hinton, and J.~Dean, ``Outrageously large neural networks: The sparsely-gated mixture-of-experts layer,'' in \emph{International Conference on Learning Representations}, 2017.

\bibitem{su2024roformer}
J.~Su, M.~Ahmed, Y.~Lu, S.~Pan, W.~Bo, and Y.~Liu, ``Roformer: Enhanced transformer with rotary position embedding,'' \emph{Neurocomputing}, vol. 568, p. 127063, 2024.

\bibitem{taori2023stanford}
R.~Taori, I.~Gulrajani, T.~Zhang, Y.~Dubois, X.~Li, C.~Guestrin, P.~Liang, and T.~B. Hashimoto, ``Stanford alpaca: An instruction-following llama model,'' 2023.

\bibitem{thambiraja2012survey}
E.~Thambiraja, G.~Ramesh, and D.~R. Umarani, ``A survey on various most common encryption techniques,'' \emph{International Journal of Advanced Research in Computer Science and Software Engineering}, 2012.

\bibitem{touvron2023llama}
H.~Touvron, T.~Lavril, G.~Izacard, X.~Martinet, M.-A. Lachaux, T.~Lacroix, B.~Rozi{\`e}re, N.~Goyal, E.~Hambro \emph{et~al.}, ``Llama: Open and efficient foundation language models,'' \emph{arXiv preprint arXiv:2302.13971}, 2023.

\bibitem{vaswani2017attention}
A.~Vaswani, N.~Shazeer, N.~Parmar, J.~Uszkoreit, L.~Jones, A.~N. Gomez, {\L}.~Kaiser, and I.~Polosukhin, ``Attention is all you need,'' in \emph{Annual Conference on Neural Information Processing Systems}, vol.~30, 2017.

\bibitem{wan2024information}
Z.~Wan, A.~Cheng, Y.~Wang, and L.~Wang, ``Information leakage from embedding in large language models,'' \emph{arXiv preprint arXiv:2405.11916}, 2024.

\bibitem{wan2024efficient}
Z.~Wan, X.~Wang, C.~Liu, S.~Alam, Y.~Zheng, J.~Liu, Z.~Qu, S.~Yan, Y.~Zhu, Q.~Zhang \emph{et~al.}, ``Efficient large language models: A survey,'' \emph{Transactions on Machine Learning Research}, 2024.

\bibitem{wu2025know}
G.~Wu, Z.~Zhang, Y.~Zhang, W.~Wang, J.~Niu, Y.~Wu, and Y.~Zhang, ``I know what you asked: Prompt leakage via kv-cache sharing in multi-tenant llm serving,'' in \emph{Network and Distributed System Security Symposium}, 2025.

\bibitem{yang2024qwen2}
A.~Yang, B.~Zhang, B.~Hui, B.~Gao, B.~Yu, C.~Li, D.~Liu, J.~Tu, J.~Zhou, J.~Lin \emph{et~al.}, ``Qwen2.5-math technical report: Toward mathematical expert model via self-improvement,'' \emph{arXiv preprint arXiv:2409.12122}, 2024.

\bibitem{yang2024first}
H.~Yang, D.~Zhang, Y.~Zhao, Y.~Li, and Y.~Liu, ``A first look at efficient and secure on-device llm inference against kv leakage,'' in \emph{Workshop on Mobility in the Evolving Internet Architecture}, 2024, pp. 13--18.

\bibitem{yu2019differentially}
L.~Yu, L.~Liu, C.~Pu, M.~E. Gursoy, and S.~Truex, ``Differentially private model publishing for deep learning,'' in \emph{IEEE Symposium on Security and Privacy}.\hskip 1em plus 0.5em minus 0.4em\relax IEEE, 2019, pp. 332--349.

\bibitem{yuan2025scx}
M.~Yuan, L.~Zhang, L.~Zeng, S.~Jiang, B.~Yang, D.~Duan, and G.~Xing, ``Scx: Stateless kv-cache encoding for cloud-scale confidential transformer serving,'' in \emph{Proceedings of the ACM SIGCOMM 2025 Conference}, 2025, pp. 39--54.

\bibitem{zhang2022opt}
S.~Zhang, S.~Roller, N.~Goyal, M.~Artetxe, M.~Chen, S.~Chen, C.~Dewan, M.~Diab, X.~Li, X.~V. Lin \emph{et~al.}, ``Opt: Open pre-trained transformer language models,'' \emph{arXiv preprint arXiv:2205.01068}, 2022.

\bibitem{zhang2020bertscoreevaluatingtextgeneration}
T.~Zhang, V.~Kishore, F.~Wu, K.~Q. Weinberger, and Y.~Artzi, ``{BERTScore}: Evaluating text generation with {BERT},'' in \emph{International Conference on Learning Representations}, 2020.

\bibitem{zhang2023h2o}
Z.~Zhang, Y.~Sheng, T.~Zhou, T.~Chen, L.~Zheng, R.~Cai, Z.~Song, Y.~Tian, C.~Ré, C.~Barrett, Z.~Wang, and B.~Chen, ``H$_2$o: Heavy-hitter oracle for efficient generative inference of large language models,'' \emph{Advances in neural information processing systems}, vol.~36, pp. 34\,661--34\,710, 2023.

\bibitem{zhao2023survey}
W.~X. Zhao, K.~Zhou, J.~Li, T.~Tang, X.~Wang, Y.~Hou, Y.~Min, B.~Zhang, J.~Zhang, Z.~Dong \emph{et~al.}, ``A survey of large language models,'' \emph{arXiv preprint arXiv:2303.18223}, vol.~1, no.~2, 2023.

\bibitem{zheng2023lmsyschat1m}
L.~Zheng, W.-L. Chiang, Y.~Sheng, T.~Li, S.~Zhuang, Z.~Wu, Y.~Zhuang, Z.~Li, Z.~Lin, E.~P. Xing, J.~E. Gonzalez, I.~Stoica, and H.~Zhang, ``Lmsys-chat-1m: A large-scale real-world llm conversation dataset,'' in \emph{International Conference on Learning Representations}, 2024.

\bibitem{zhou2024survey}
Z.~Zhou, X.~Ning, K.~Hong, T.~Fu, J.~Xu, S.~Li, Y.~Lou, L.~Wang, Z.~Yuan, X.~Li \emph{et~al.}, ``A survey on efficient inference for large language models,'' \emph{arXiv preprint arXiv:2404.14294}, 2024.

\end{thebibliography}
\end{document}